\documentclass[pre,twocolumn,floatfix,showpacs]{revtex4}
\usepackage{color}
\usepackage{graphicx}
\usepackage{amsmath}
\usepackage{array}

\providecommand{\tabularnewline}{\\}

\newcommand{\velo}{\upsilon}
\newcommand{\erfc}{\mathrm{erfc}}

\newcommand{\bibfnamefont}[1]{#1}
\newcommand{\bibnamefont}[1]{#1}
\newcommand{\citenamefont}[1]{#1}

\begin{document}
\title{Fracture as a pattern formation process}
\author{M. Fleck}
\email{michael.fleck@uni-bayreuth.de}
\author{D. Pilipenko}
\affiliation{Lehrstuhl f\"ur Material und Prozesssimulation, Universit\"at Bayreuth, Germany}
\author{R. Spatschek}
\affiliation{Max-Planck-Institut f\"ur Eisenforschung GmbH, D\"usseldorf, Germany}
\author{E. A. Brener}
\affiliation{Institut f\"ur Festk\"orperforschung, Forschungszentrum J\"ulich, Germany}

\pacs{62.20.mm, 46.15.-x, 46.50.+a, 47.54.-r}

\date{\today}

\begin{abstract}
A continuum model of crack propagation is presented and discussed.
We obtain steady state solutions with a self-consistently selected propagation velocity and shape of the crack, provided that elastodynamic and viscoelastic effects are taken into account.
Two different mechanism of crack propagation, a first order phase transition and surface diffusion are considered, and we discuss different loading modes.
The arising free boundary problems are solved by phase field methods and a sharp interface approach using a multipole expansion technique.
\end{abstract}

\maketitle

\section{Introduction}

The dynamics of crack propagation is an important and long standing mystery in solid-state physics and materials science \cite{Lawn:1993fp, marder:24}, and in the recent years the physics community experienced a rebirth of interest in the problem of dynamic fracture.
The fundamental basis of today's understanding of the phenomenon fracture traces back to Griffith \cite{Griffit1921}, who realized that the growth of cracks is determined by a competition of a release of elastic energy and a simultaneous increase of the surface energy due to the advancing crack. 
The uniform motion of a crack is relatively well understood in the framework of continuum theories \cite{Freund:1998yq, Broberg:1999vn, Rice:1968kx}.
Here, the conventional approach is to treat the  crack as a front or interface separating broken and unbroken regions of the material;
propagation is governed by the balance of the elastic forces in the materials and cohesive stresses near the crack tip \cite{PhysRevLett.76.1497, PhysRevLett.82.2314, PhysRevLett.79.877}.
Many characteristic features of crack propagation are nowadays well established by experimental studies \cite{PhysRevLett.82.3823, PhysRevLett.74.5096, PhysRevLett.76.2117, PhysRevLett.67.457, PhysRevE.53.5637, Sharon:1999ly, 0295-5075-30-6-004, BoudetCilibertoSteinberg1996}.
As soon as the flux of energy to the crack tip exceeds a critical value, the crack becomes unstable and starts to branch while emitting sound waves.
These phenomena are consistent with the continuum theory of sharp crack tips, but it fails to describe them, because the details of crack growth, in particular the chosen crack path and velocity, depend on details of cohesion on microscopic scales \cite{Fineberg:1999zr}.
Nevertheless, empirical energy balances and simple propagation laws that are frequently used in engineering applications, cannot account for the richness of actual fracture phenomena. In particular, they cannot predict
dynamical instabilities of fast moving cracks. The fundamental mechanisms of these instabilities have been extremely difficult to elucidate because they appear to result from a non-trivial coupling between dynamical phenomena inside the crack tip region, known as {\em process zone}, and (linear) elasticity, with no clear separation of scale between atoms and the system size. 
 
Large scale molecular dynamics (MD) simulations with about $10^7$ atoms allowed to get deeper insights into the growth behavior of cracks \cite{PhysRevLett.77.869, PhysRevLett.78.479, PhysRevLett.78.689, PhysRevLett.79.1309}.
Although limited to submicron samples and very short timescales, these simulations were able to reproduce key features of crack propagation like the initial acceleration and the onset of instabilities.
Nevertheless, a detailed understanding of the complex physics of crack propagation, in particular aspects of the pattern formation process, still remain a major challenge \cite{AlexanderHellemans08141998}.

At this level, continuum descriptions, in particular phase field methods that avoid dynamical artifacts which are associated with the breaking of translational and rotational symmetry \cite{PhysRevE.68.036118, Marder19951}, offer a useful and complementary perspective on crack propagation as a pattern formation process.
The past years have seen intense activities in phase field modeling of crack propagation (see \cite{SpatschBrenerKarma012010} for a recent review) and of defects in general \cite{WangDefect2010}.

Here, we propose a continuum description of crack propagation in the spirit of interfacial pattern formation processes. 
Inspired by the discovery of the Asaro-Tiller-Grinfeld (ATG) instability  \cite{AsaroTiller,Grinfeld86,Srolovi1989}, we understand fracture as late and highly nonlinear stage of this elastically driven interfacial instability. 
In its early stage, a linear stability analysis of a solid surface under uniaxial load reveals that long wave morphological perturbations are unstable in the sense that they lead to a decrease of the total free energy. 
Finally, in a later stage of the instability one observes the formation of deep notches, which are similar to cracks (see e.g.~\cite{YangSrolovi091993,SpencerMeiron111994,KassnerMisbah1994}).
Nevertheless, if solely accounting for linear elasticity, this instability leads to a breakdown of the physical description due to a finite time cusp singularity\index{finite time cusp singularity}:
After a finite time, the unstable deep grooves advance with infinitely high velocities and vanishing tip radii (see e.g.~\cite{LoPomyalovProcaccia2008}).

This problem can be solved, for instance, by the inclusion of elastodynamic effects which restore the selection of the steady state tip radius and velocity. 
Based on this recognition, a minimal continuum theory of fracture was developed using only well-established
thermodynamical concepts \cite{BrenerSpatch2003}. 
In this picture, a full modeling of a propagating crack not only determines the crack speed but also the entire crack shape and scale self-consistently, which leads to a description as a moving
boundary problem. 
The latter was then solved by basically two different methods: 
First, a sharp interface multipole expansion technique \cite{pilipenko2007} and the fully dynamic phase field method \cite{Spatchek2006,Spatschek2007}. 
Remarkably, already this single parameter minimum model selects steady state propagation velocities appreciably below the Rayleigh speed and shows a tip splitting instability for high applied tensions. 
A shortcoming of the model is a decaying velocity as function of the driving force over a significant range of applied stresses.

Recently, a similar continuum model of fracture was proposed in \cite{SpatschekBrenerPilipenko2008}, curing the problem of the finite time singularity by 
viscoelastic damping.
Apart from the usual dissipation directly at the crack surface, viscous bulk dissipation takes place in an extended zone around the crack.
Hence, the incoming flow of elastic energy is partially converted to surface energy, in order to advance the crack, and thermal energy due to viscous damping.
However, this model does not capture a branching instability for high applied loads in case of mode I loading.

The purpose of the present paper is twofold:
First, it summarizes and extends the aforementioned work for the limiting cases of pure elastodynamics and viscoelasticity.
Second, it introduces a description which contains {\em both} effects, therefore capturing the benefits of them and overcoming the limitations.
We apply the model mainly to mode I, but also mixed mode loadings consisting both of mode I and mode III.
Two different material transport mechanisms are considered and compared.


The paper is organized as follows: In section \ref{sec:crack-model}
we present the continuum model of crack propagation in elastodynamic and viscoelastic
media. 
Then, in section \ref{selection}, the crack tip selection principles are discussed. 
The arising free boundary problems are solved numerically by the use of sharp interface and phase field methods, as presented in section \ref{methods}. Finally,  we discuss the predictions
of the model in section \ref{resultssec}.

\section{Continuum Model of Fracture}
\label{sec:crack-model}

\begin{figure}
\begin{center}
a)
\includegraphics[trim=3.5cm 4cm 3.5cm 2cm, clip=true, width=3.5cm]{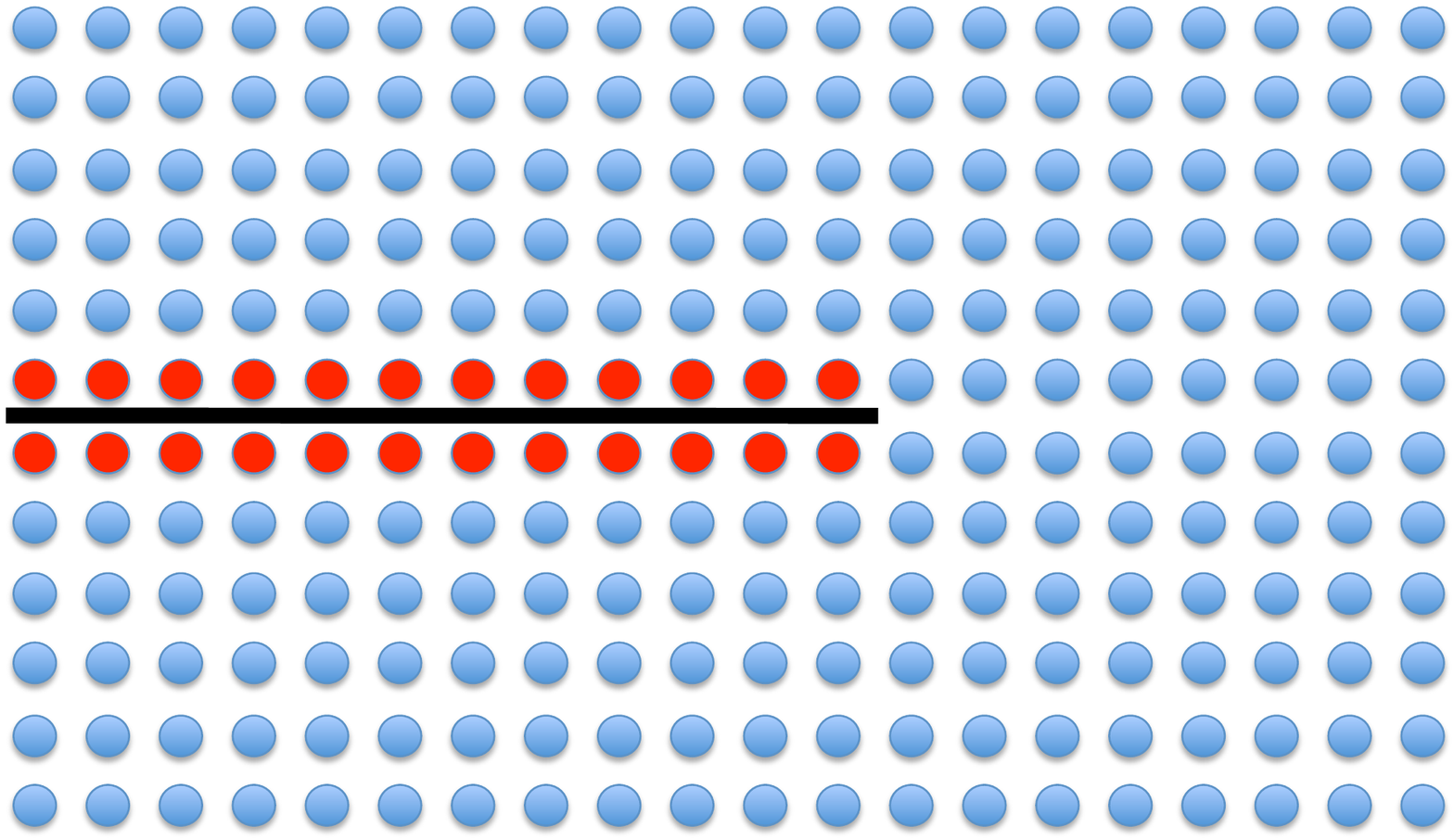}
b)
\includegraphics[trim=3.5cm 4cm 3.5cm 2cm, clip=true, width=3.5cm]{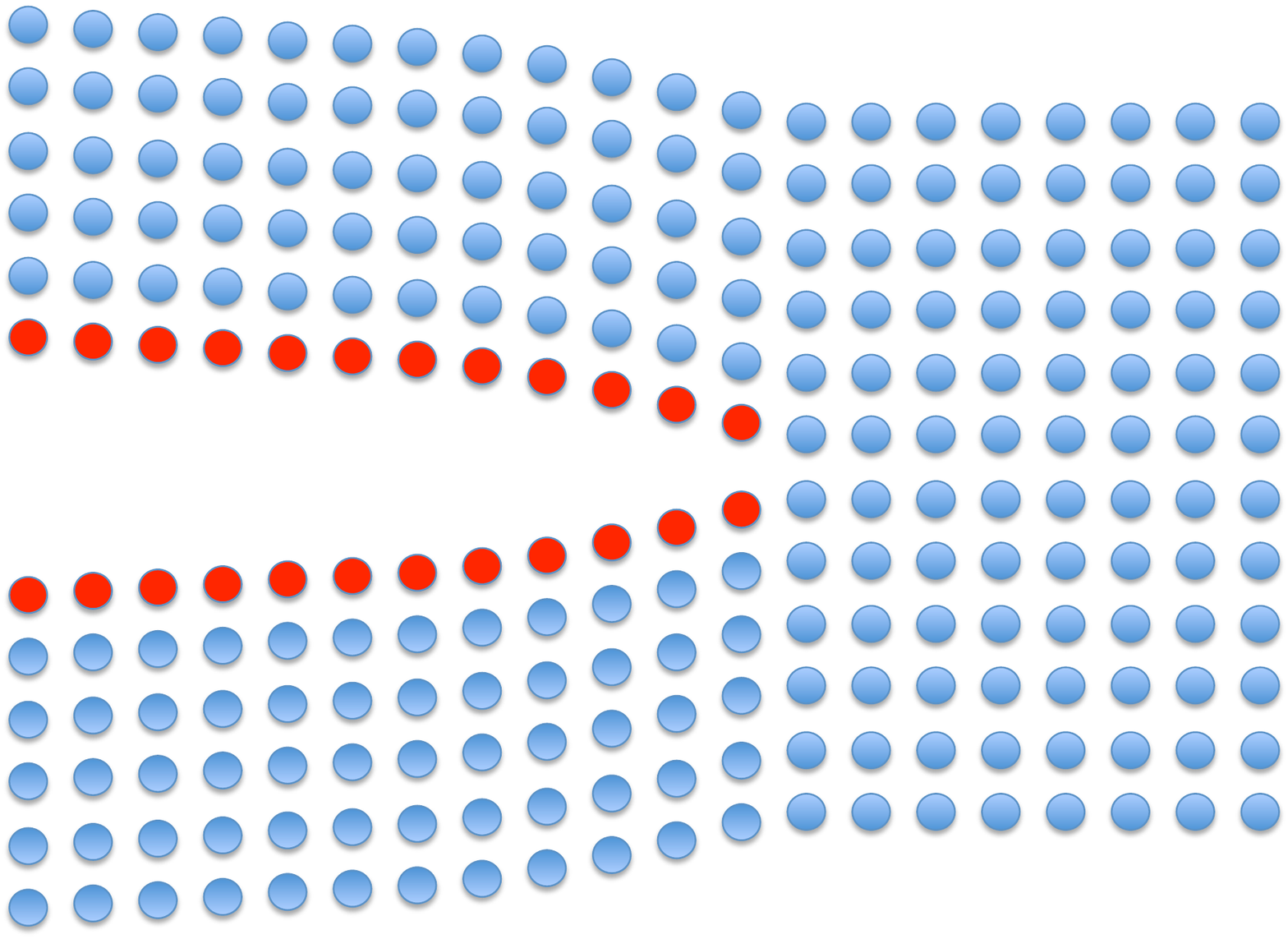}

c)
\includegraphics[trim=3.5cm 4cm 3.5cm 2cm, clip=true, width=3.5cm]{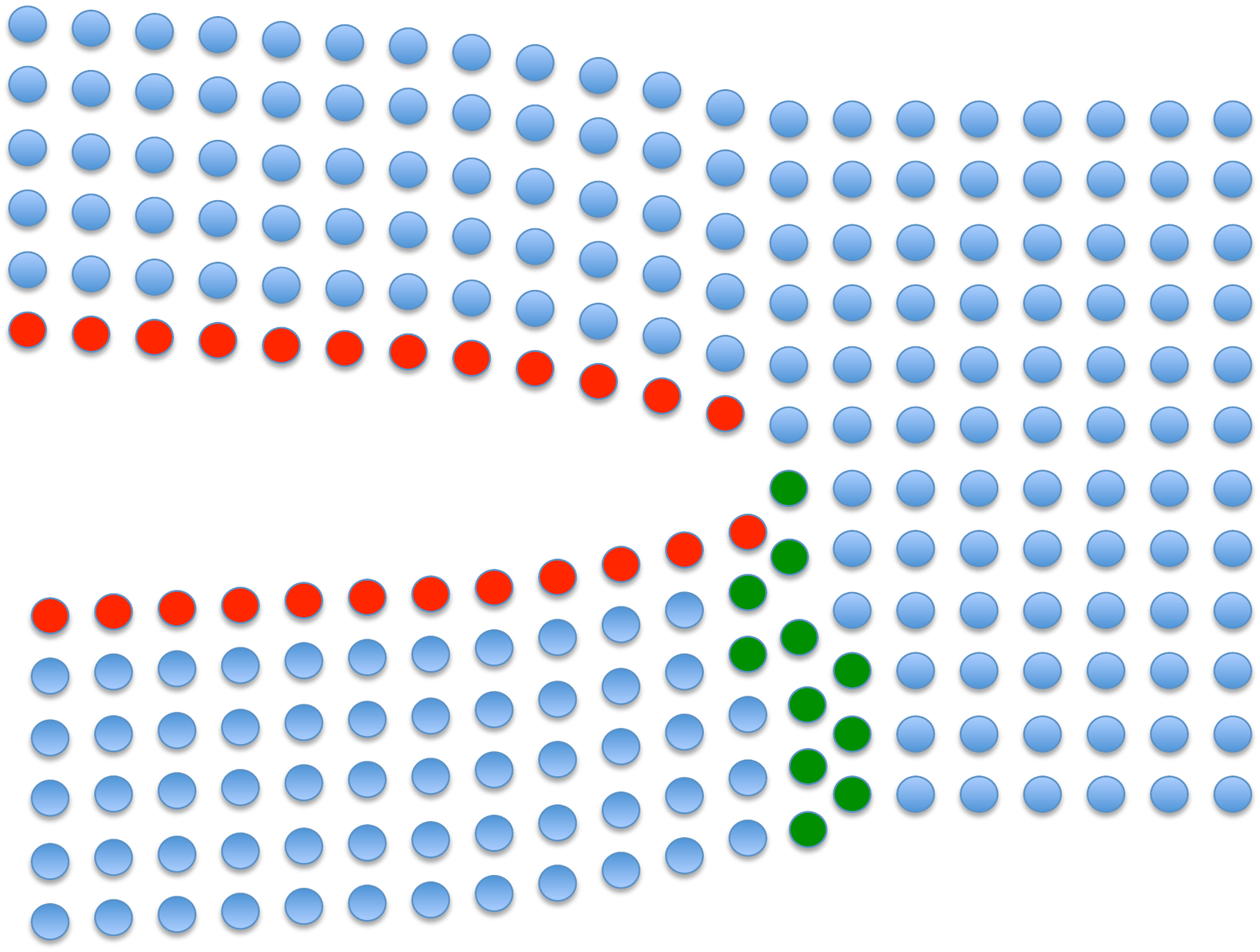}

d)
\includegraphics[trim=3.5cm 4cm 3.5cm 2cm, clip=true, width=3.5cm]{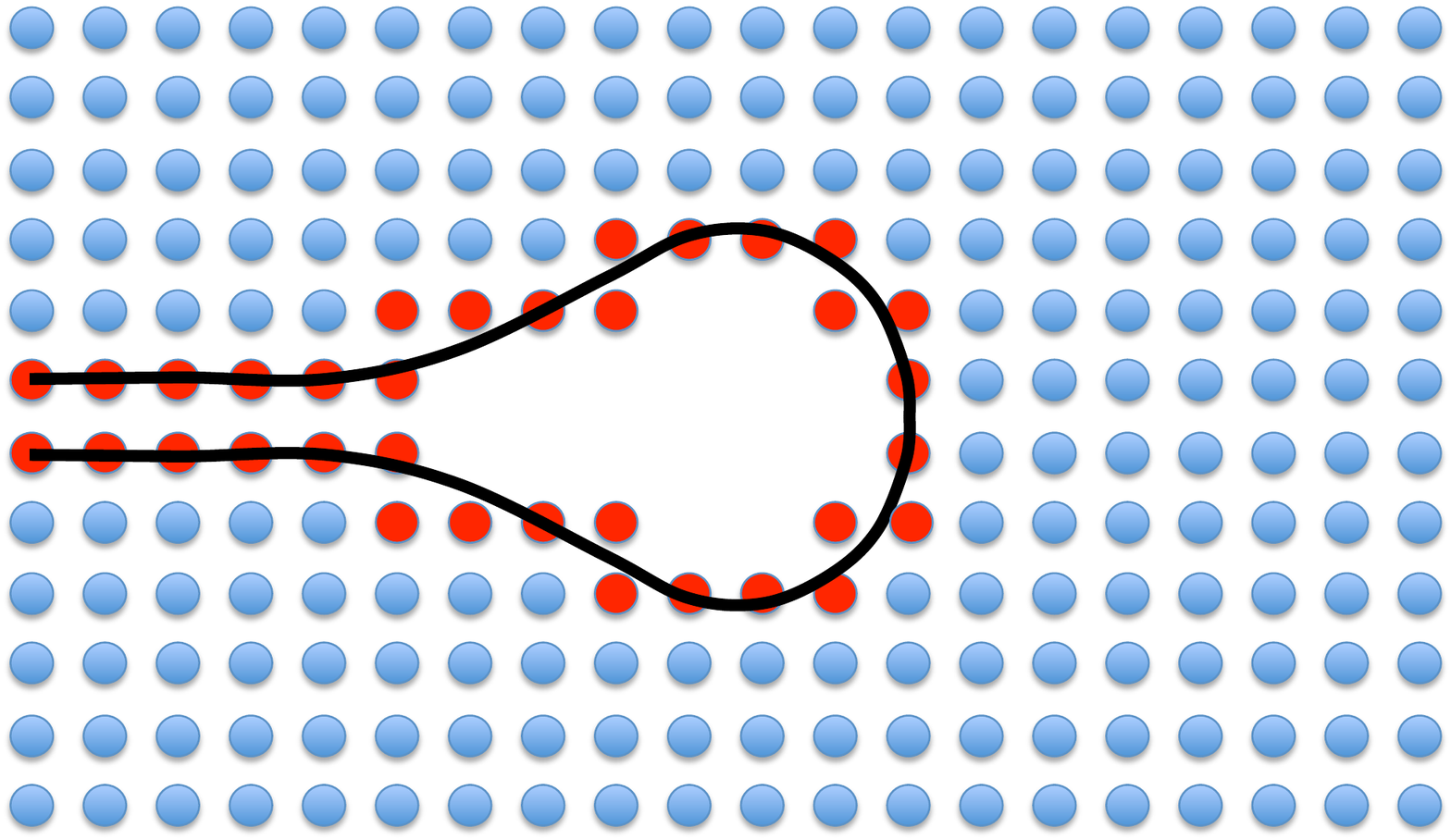}
e)
\includegraphics[trim=3.5cm 4cm 3.5cm 2cm, clip=true, width=3.5cm]{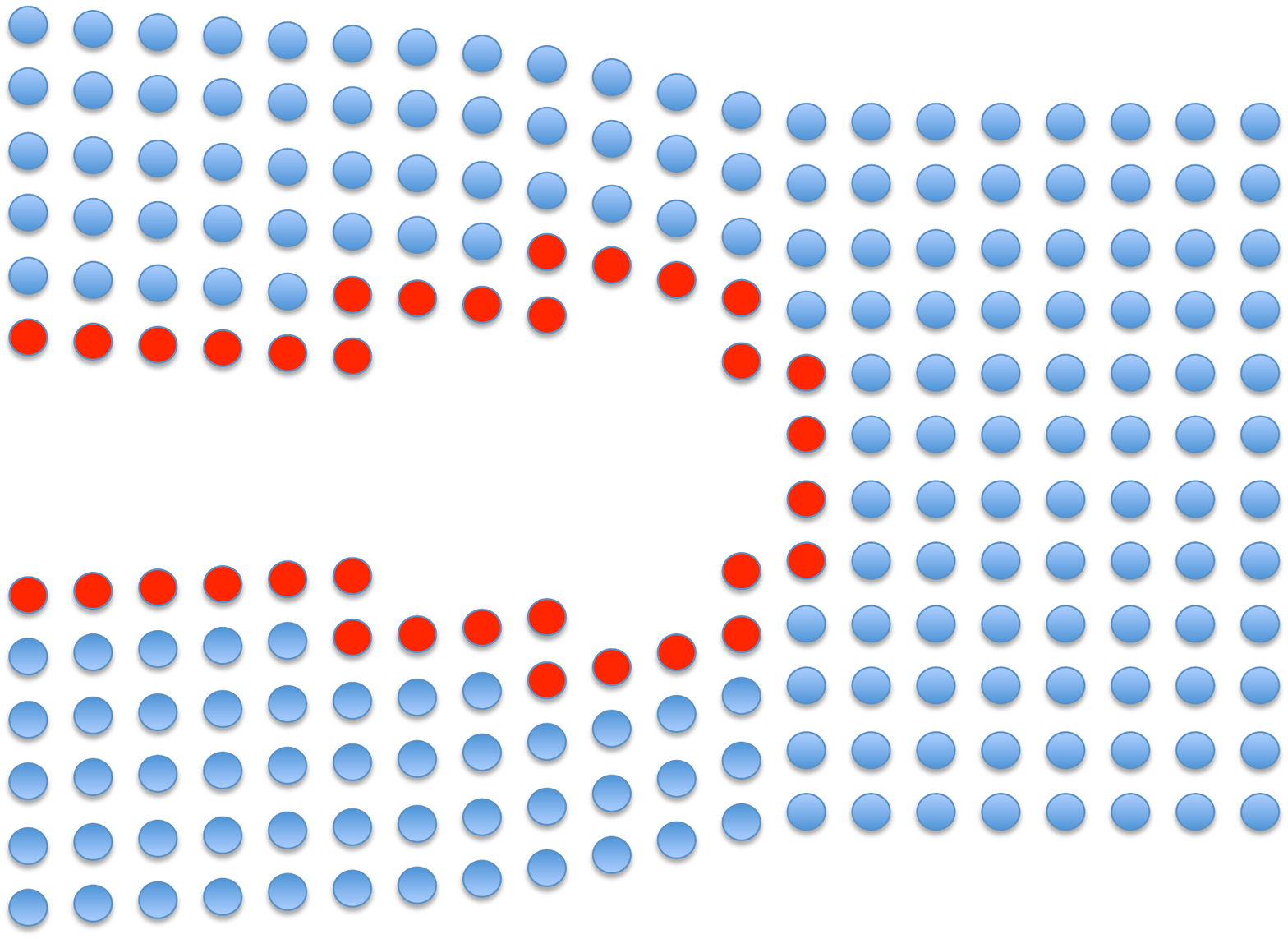}

f)
\includegraphics[trim=3.5cm 4cm 3.5cm 2cm, clip=true, width=7.0cm]{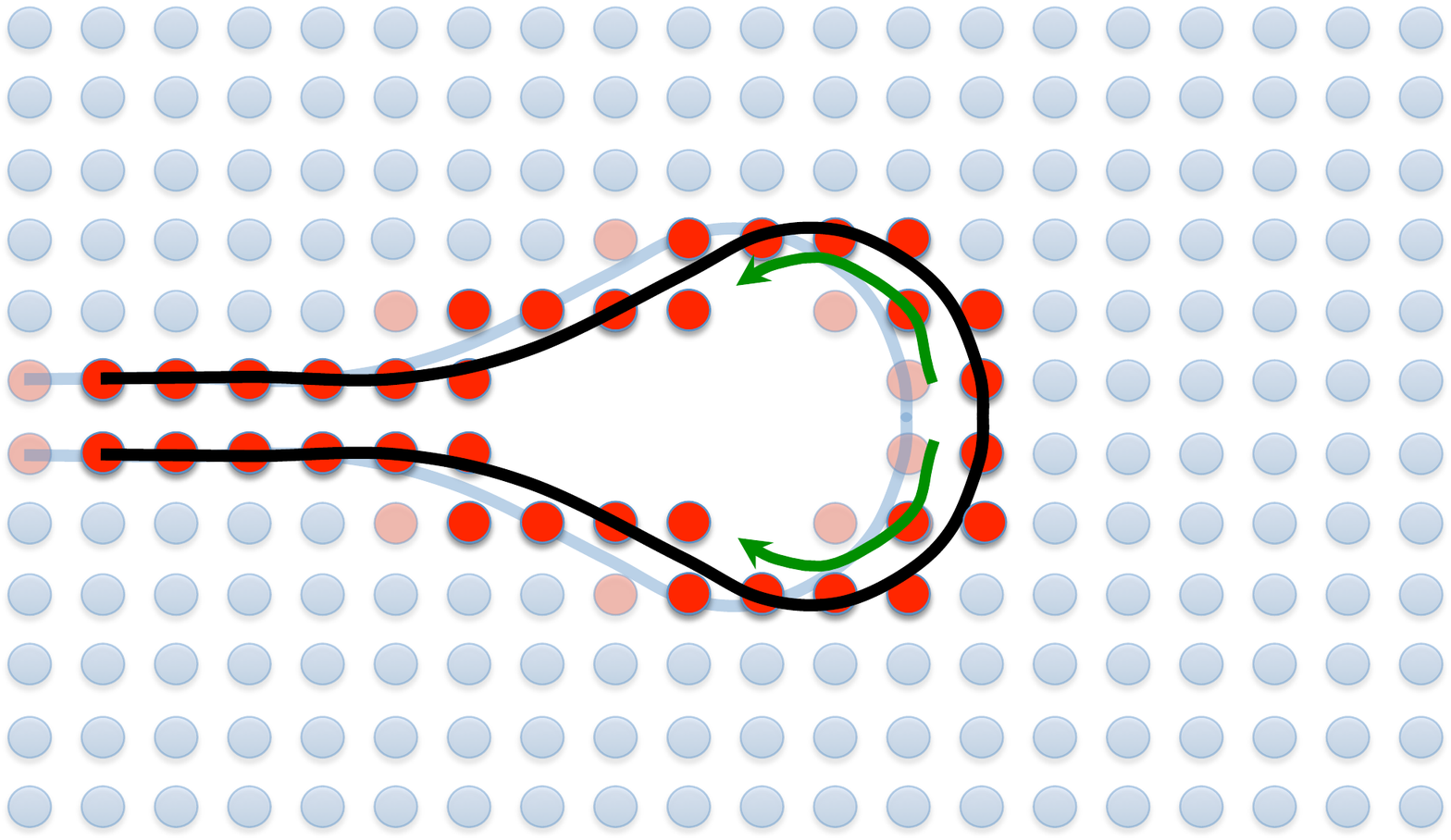}
\caption{Sketch of crack propagation mechanisms. (a) Bond breaking picture, depicted in the reference frame. The black line indicates the crack, and there mechanical boundary conditions have to be applied. The atoms which change their atomic configuration due to bond breaking are shown in red. (b) The same crack in the deformed configuration, showing the separation of interface atoms. (c) Dislocation emission from the crack tip leads to blunting. The atomic neighborhood relations change in the bulk, as illustrated by the green atoms. (d) Sketch of the atomic configuration in the reference frame for a crack with finite tip radius. (e) The same in the deformed state. (f) Propagation of a crack with finite tip radius demands mass transport, since atoms have to be removed from the tip region. Here we illustrate the crack growth by surface diffusion. The transparent background shows the configuration in the previous timestep, the solid red atoms the interface atoms after advance by one lattice unit.}
\label{atomfig}
\end{center}
\end{figure}

We propose a description of fracture in the spirit of elastically
driven interfacial pattern formation processes. In contrast to classical
descriptions, where the tip is treated as a singular point followed
by a mathematical cut, we assume the crack to be macroscopically extended,
and, even more important, to have a finite tip radius $r_{0}\sim h$,
as can be seen in Figs.~\ref{atomfig}, \ref{GeometrySD} and \ref{GeometryPT}.

This finite tip size implies, that the volume inside the crack is also finite, and -- depending on the growth mechanism of the crack -- also a description of this inner ``phase'' is necessary.
The shape of the crack itself is not an input to the model, but is determined self-consistently by the equations of motion for the {\em entire} crack.
In this sense, the description differs substantially from classical models, where only equations of motion for the singular crack tip have to be postulated or derived.
One of the advantages of such a description is that the entire crack shape is a degree of freedom for the model, and therefore not only the advance of the crack itself is described, but also deformations of the crack contour behind the tip, and -- what is much more important -- path selection is automatically contained in such models \cite{Hakim}.

The equations of motion for the crack depend on the local elastic deformations, but also the local curvature of the interface.
In this way, they naturally capture the effect of stress release through crack propagation, but also the increase of interfacial energy due to crack elongation, which is the basis for the Griffith criterion.
It turns out that the desired self-consistent selection of the crack shape is a nontrivial step, since by the aforementioned ATG instability the tip tends to become sharper and sharper, without a self-consistent selection of a tip scale, if only static linear elasticity and interfacial energy are taken into account.

Another important aspect is related to the definition of the crack shape.
We describe all patterns here in the Lagrangian reference frame, which means that in this configuration the mechanical deformations are {\em not} taken into account.
In this reference frame a straight cut, which is frequently used for a mathematical description of a crack, would just appear as a line, and has a vanishing tip radius.
Under deformation, however, the crack surfaces separate, in particular the distance between the lips would scale as $\Delta u \sim K_I r^{1/2}/E$, where $r$ is the distance from the crack tip, $K_I$ the mode I stress intensity factor and $E$ is the elastic modulus of the material.
Pure elasticity describes only the deformation of materials, but it does not provide evolution equations of the motion of the crack. 
In particular, {\em linear} elasticity would predict a $\sigma\sim K_I r^{-1/2}$ stress singularity at an infinitely sharp crack tip.
Physically, one would expect regularization of this singularity either by nonlinear effects or a finite tip radius $r_0$, which serves as a cutoff parameter.
In this work we do not follow the first regularization approach and consider only linear mechanical models, which will be linear elastodynamics and a linear (Kelvin) viscoelastic model;
instead, we consider situations with a finite tip radius.

Let us briefly contrast this description with conventional pictures of crack growth on an atomistic level.
In brittle materials, the intuitive interpretation of crack propagation is due to the breaking of bonds at sharp crack tips.
This changes the neighborhood configuration for an atom at the crack tip only in the sense that connections to some of the adjacent atoms is lost, and simultaneously the distances to the other atoms are changed due to elastic deformations.
This breaking of bonds corresponds to the advance of the sharp cut in a continuum model (see Fig.~\ref{atomfig}a and b).
Ductile effects leads to the emission of dislocations from the crack tip, and lead to blunting (see Fig.~\ref{atomfig}c).
These plastic events introduce also configurational changes in the bulk in the sense that the neighborhood relations are changed.
We point out, that this is a bulk effect, which of course also reaches the crack, since the dislocation lines have to terminate at surfaces.
Notice that on the continuum level such modeling requires either equations of motions for dislocations or -- in a coarse grained framework -- plasticity models.
In this article, we will focus on yet another effect, which is not captured by the above picture, which is due to material rearrangement at surfaces (see Fig.~\ref{atomfig}d-f).
It means, literally speaking, that atoms are removed and attached to the crack surfaces at different places, and therefore the neighborhood configurations are changed now in the sense of a surface effect.
For a crack that has a finite tip radius already in the reference frame, advance of the crack means that atoms have to be removed from the tip.
They can be deposited again on different regions of the crack surface, and in this case we can interpret the material transport as a surface diffusion process.
Alternatively, we could imagine that the removed atoms become part of a ``gas phase'' inside the crack.
A gas has of course a lower density than the solid, which would require that ultimately the gas atoms have to be ejected from the crack (if the density difference is not accommodated by the crack opening).
For convenience, we will not consider this fast ``hydrodynamic'' transport and ignore the density difference between the solid and the ``dense gas phase''.
Notice that in both cases of material transport the atoms undergo long-range transport (on the atomic scale), and therefore their neighborhood configuration changes completely.

On the continuum level, we therefore have to provide equations of motion for each interface point of the crack, reflecting either surface diffusion (SD) or the phase transformation (PT) mechanism between the solid and the gas phase.
They are coupled to the elastic fields in a nonlocal and nonlinear manner.
The motion is then driven by the tendency to lower the total free energy of the system.
An important and obvious difference is that for SD the number of ``solid'' atoms is conserved, which is not the case for the PT mechanism.


For both transport mechanisms, we consider the growth of a single crack in a strip geometry, in order to have a constant stress intensity factor. 
We restrict to an effectively two dimensional system by the assumption of translational invariance
in the $z$-direction, and assume the strip to be infinitely extended in the direction of propagation, which in our case is chosen to be the $x$-direction. 
We mainly concentrate on mode I fracture, which means that the applied tensile forces act
in the $y$-direction perpendicularly to the crack faces. 
Apart from this, we will also discuss results from the application of mode III loadings, and linear combinations of these two modes.
Since the crack tip is macroscopically extended, no singularity appears and the whole crack pattern can  be described consistently in the continuum approximation.

In a Lagrangian description of linear elasticity, the elastic state of the system is described through a continuous displacement field $u_{i}$.
Then, the strains are defined as the symmetrical spatial derivatives of the displacements, 
\begin{align} 
\label{mod::eq1}
\epsilon_{ik} & =\frac{1}{2}\left(\frac{\partial u_{i}}{\partial x_{k}}+\frac{\partial u_{k}}{\partial x_{i}}\right).
\end{align}
As the total stress field depends linearly on both the strain as well as the strain--rate, we conveniently decompose it into a strain and a strain--rate dependent part, 
\begin{align}
\sigma_{ik}^{(tot)} & =\sigma_{ik}^{(el)}(\epsilon_{ik})+\sigma_{ik}^{(vis)}(\dot{\epsilon}_{ik}),\end{align}
where $\sigma_{ik}^{(el)}$ and $\sigma_{ik}^{(vis)}$ are the elastic
and viscous stresses, respectively, and $\dot{\epsilon}_{ik}$ denotes the time derivative of the strain ${\epsilon}_{ik}$. 
Furthermore, we restrict the considerations to fully isotropic media. 
Then, as given by Hooke's law, the elastic stresses are 
\begin{align}
\sigma_{ik}^{(el)} & =\frac{E}{1+\nu}\left(\epsilon_{ik}+\frac{1}{1-2\nu}\,\delta_{ik}\epsilon_{ll}\right),\label{HookeLaw}
\end{align}
where $E$ is Young's modulus and $\nu$ the Poisson ratio, and we use the Einstein sum convention.
By construction, the viscous stresses are formally similar to the elastic stresses \cite{Landau::Elasticity}, and we therefore write them for a Kelvin viscoelasticity model as 
\begin{align}
\sigma_{ik}^{(vis)} & =\frac{\eta}{1+\zeta}\left(\dot{\epsilon}_{ik}+\frac{1}{1-2\zeta}\,\delta_{ik}\dot{\epsilon}_{ll}\right),
\label{ViscousStresses}
\end{align}
with the two viscous constants $\eta$ and $\zeta$.

The evolution of the elastic degrees of freedom within the viscoelastic solid is given by Newton's equation of motion, and the elastic displacements $u_{i}$ have to fulfill 
\begin{align}
\frac{\partial\sigma_{ik}^{(tot)}}{\partial x_{k}} & =\rho\ddot{u}_{i},
\label{bulk}
\end{align}
where $\rho$ is the mass density. 
This equation ensures locally a force balance between the elastic stress and viscous friction on
the left hand side and inertia on the right side.
These equations have to be supplemented by mechanical boundary conditions, which are given below for the two different transport mechanisms.

\subsection{Surface diffusion}

\begin{figure}
\begin{center}
\includegraphics[angle=0,width=0.9\linewidth]{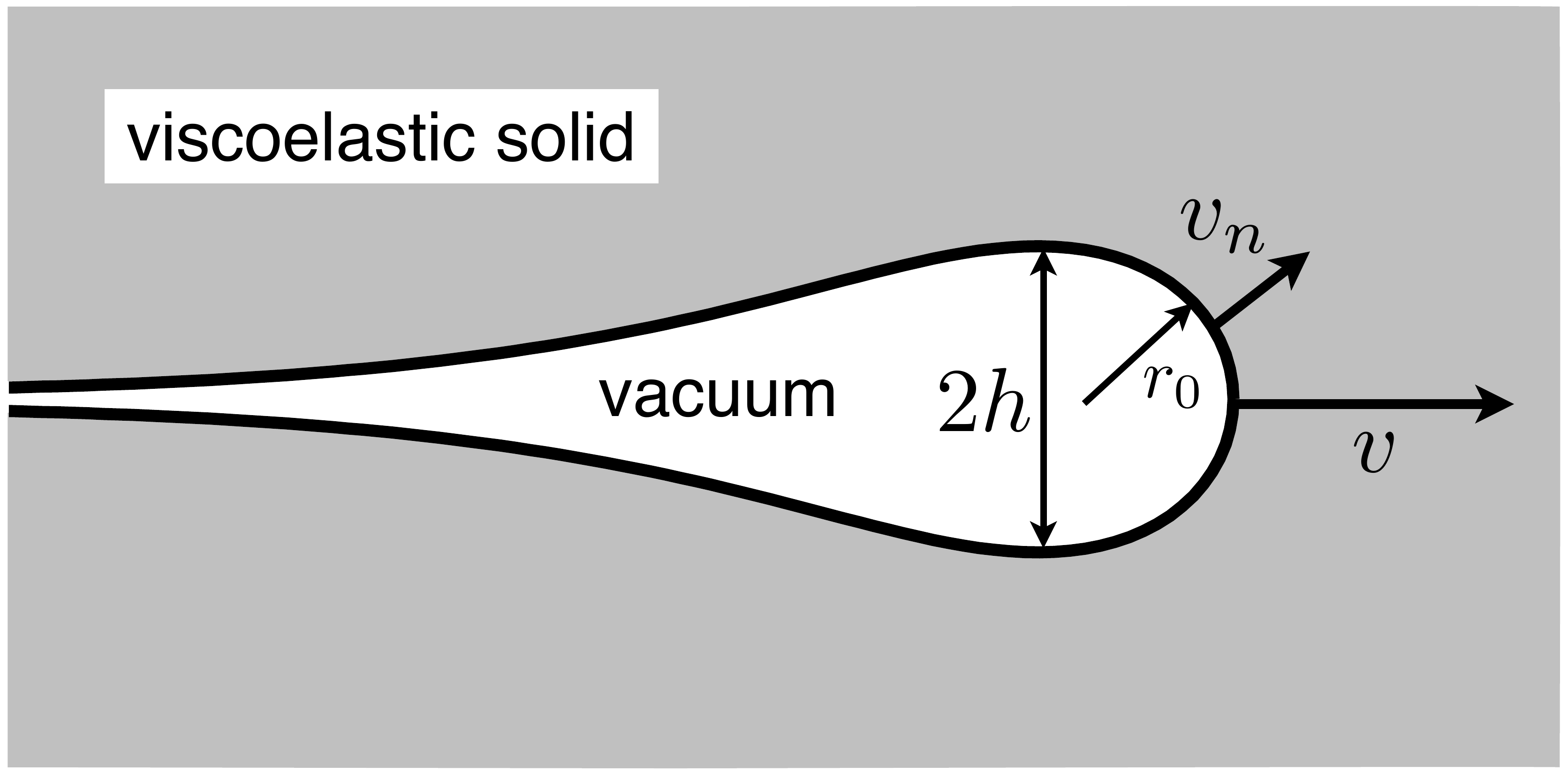}
\caption{Schematic picture of the steady state crack propagation by surface
diffusion. The crack contour, indicated by the solid black line, separates
the viscoelastic medium from the advancing ``vacuum bubble''.
During
the propagation the total amount of solid material is conserved.
}
\label{GeometrySD}
\end{center}
\end{figure}

For crack growth by surface diffusion, the crack is filled with vacuum, and therefore we impose stress free boundary conditions at the crack contour, 
\begin{align}
\sigma_{in}+\rho\velo_{n}\dot{u}_{i} & =0,
\label{MechanicalBoundCond}
\end{align}
where $n$ is the direction normal to the interface, and $\velo_{n}$ is the normal interface velocity (see Fig.~\ref{GeometrySD}). 
The second term on the left hand side accounts for momentum conservation at the solid--vacuum interface. 
We point out, that in the dynamic limit, when the crack propagation velocity $\velo$ is of the order of the materials sound speed, this term becomes important \cite{Freund:1998yq}.

So far, for an arbitrarily given crack shape and known strain history, the mechanical problem
is unambiguously determined and can be calculated by Eqs.~(\ref{mod::eq1})--(\ref{MechanicalBoundCond})
together with the outer boundary conditions at the borders of the strip, which specify the externally applied loading. 
Next, we have to formulate an evolution equation for the crack contour. 
The motion of the interface is caused by thermodynamically induced mass transport processes, which diminish the total free energy of the system.
The local driving force for crack propagation is given by the chemical potential $\mu$ at the solid vacuum interface \cite{Nozieres1993},
\begin{align}
\mu & =\Omega\left(\frac{1}{2}\sigma_{ik}^{(el)}\epsilon_{ik}-\frac{1}{2}\rho\dot{u}_{i}^{2}-\gamma\kappa\right).
\label{steady:eq1}
\end{align}
with $\gamma$ being the surface energy per unit area and $\kappa$ the surface curvature, which is counted to be positive, if the crack shape is convex; the atomic volume $\Omega$ appears since the chemical potential is defined as free energy per particle. 
We point out that the viscous stresses do not appear in the chemical potential, since viscous dissipation is a sole property of the bulk, whereas the chemical potential is needed to describe energy dissipation through the motion of the interface. 
Furthermore, we note that due to inertial effects, also the kinetic energy density appears in the chemical potential.
Counterintuitively, it appears with sign opposite to that of the potential energy; this can be derived rigorously from variational principles \cite{SpatschekFleck2007,Spatschek2007}.

For surface diffusion the motion of the crack surface is proportional
to the divergence of a flux of solid material along the interface.
This flux of material is induced by gradients of the chemical potential.
We express the motion of the interface by the local normal velocity
$\velo_{n}$ and obtain 
\begin{align}
\velo_{n} & =-\frac{D_{s}}{\gamma\Omega}\frac{\partial^{2}\mu}{\partial^{2}s},
\label{steady:eq3}
\end{align}
where $\partial/\partial s$ denotes the tangential derivative and the diffusion coefficient $D_{s}$ has a dimension $[D_{s}]=\mathrm{m}^{4}\mathrm{s}^{-1}$.
We note that for surface diffusion the amount of solid material is conserved during the crack propagation. 
A typical steady state crack shape using surface diffusion is shown in Fig.~\ref{GeometrySD}.
One can see that the crack first opens up to a tip diameter $2h$, and then closes again due to the condition of material conservation.

We point out that this description of mass transport is not limited to surface diffusion in its literal sense only.
Often, many complicated physical processes like plastic bulk flow take place in a small zone around the tip. 
Assuming that this zone is relatively thin, the mass transport can \emph{effectively} be described
by surface diffusion, where the detailed information about the process zone is hidden in the diffusion coefficient in the spirit of a lubrication approximation.

\subsection{Phase transformations}

\begin{figure}
\begin{center}
\includegraphics[angle=0,width=0.9\linewidth]{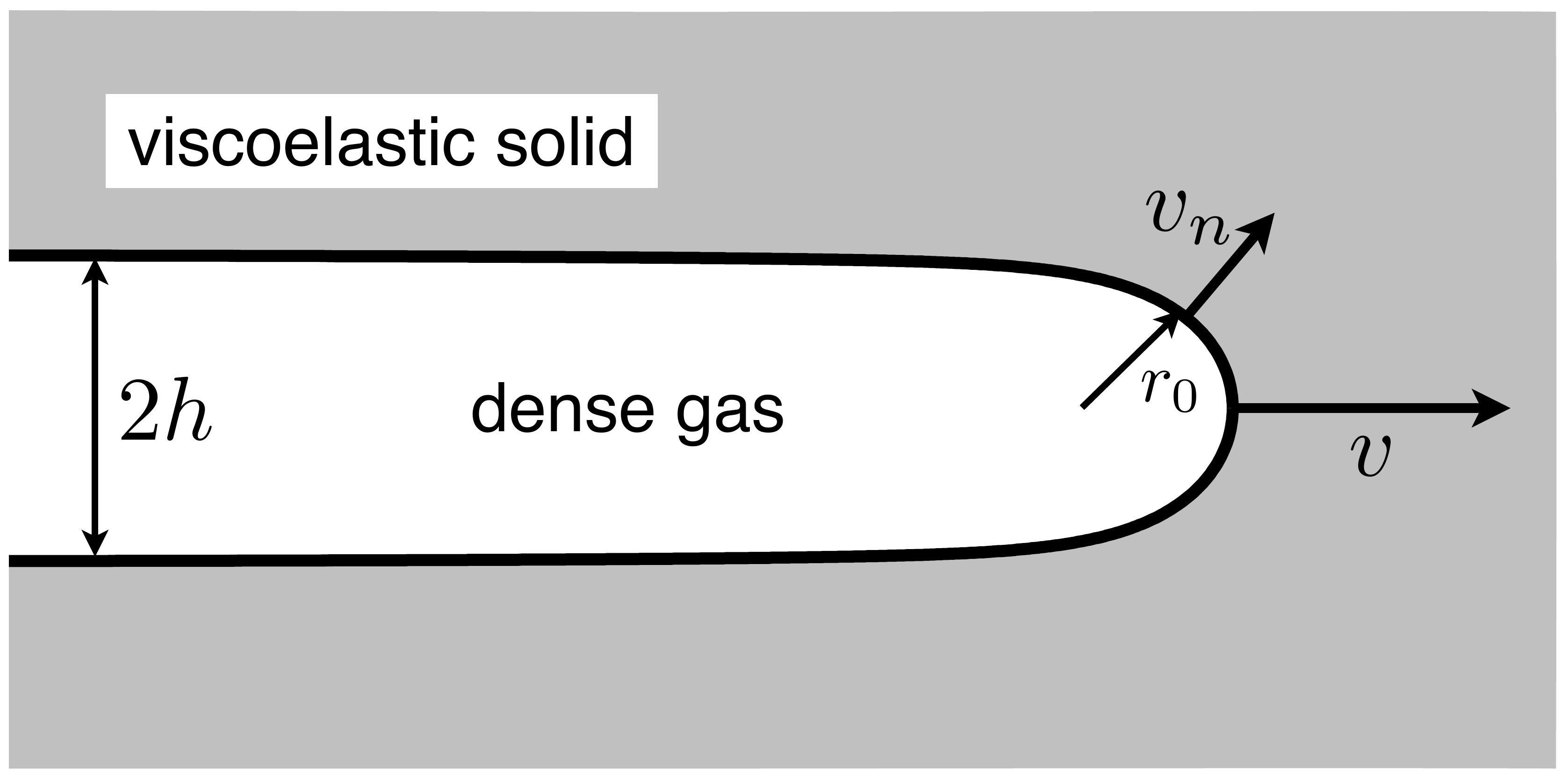} 
\caption{Sketch of a propagating crack,
where the phenomenon of fracture is interpreted as a phase transformation
process from a viscoelastic solid to a ``dense gas'' phase. The
crack surface, indicated by the solid black line, separates the original
viscoelastic medium from the growing ``dense gas'' phase. 
}
\label{GeometryPT} 
\end{center}
\end{figure}

Here we discuss crack propagation by means of a phase transformation process, where the solid matrix transforms into a ``broken gas phase'' with vanishing elastic moduli.
We assume that the gas phase and the viscoelastic medium to have equal mass densities $\rho$. Furthermore, the interface between this ``dense gas'' phase
and the medium is considered to be coherent, i.e.~the displacements
are continuous there. With these assumptions, two central simplifications
are achieved. First, instead of Eq.~(\ref{MechanicalBoundCond})
the mechanical boundary conditions are 
\begin{align}
\sigma_{in} & =0,
\label{mech_bound_cond_PT}
\end{align}
which means that no velocity dependent correction appears here, since by the continuity of velocities and densities also the momentum is contiuous.
Second, the expression for the chemical potential
is replaced by 
\begin{align}
\mu & =\Omega\left(\frac{1}{2}\sigma_{ik}^{(el)}\epsilon_{ik}-\gamma\kappa\right),\label{chemical_pot_PT}
\end{align}
 where the kinetic energy contribution does not appear. 
 The reason for this simplification is the continuity of the kinetic energy density, because the above expression of the chemical potential should be more correctly be interpreted as the chemical potential difference between the adjacent phases \cite{SpatschekFleck2007,Spatschek2007}.
 Notice that the inner phase is assumed to be infinitely soft, therefore it has a vanishing elastic energy density.
Again, the motion of the interface is locally expressed through the
normal velocity, which in this case is direct proportional to the
chemical potential difference at the interface, \begin{align}
\velo_{n} & =\frac{D}{\Omega\gamma}\mu\label{steady:eq2}\end{align}
 with a kinetic coefficient $D$ having the dimension {[}D]=m$^{2}$s$^{-1}$.
Of course, using this model, the amount of solid material is not conserved
during crack propagation. 
In this
sense, our model is strongly related to phase field models of fracture based on a non-conserved order parameter \cite{KarmaKesslerLevine2001,AraKalVin2000,HenLev2004}.
The crucial difference is that the current model is based on well-defined sharp interface equations, and therefore the predictions do not depend on inherently numerical parameters like a phase field interface width.
A typical steady state crack shape using phase transformations is shown in Fig.~\ref{GeometryPT}. In contrast to surface diffusion the crack keeps its opening, and does not close far behind the tip (in the reference state) due to the absence of material conservation.

However, although surface diffusion seems to be more adequate for a description of fracture, from a numerical point of view the treatment of Eq.~(\ref{steady:eq3}) is much more time-consuming due to the higher order spatial derivatives, which are not present in Eq.~(\ref{steady:eq2}).
Therefore, modeling of fracture as a phase transition process offers numerical advantages. 

\section{Crack propagation: Selection principles}
\label{selection} 

The self-consistent selection of the crack velocity, the tip radius and even the entire crack shape is a central aspect of the present theory, and will be discussed in more detail in this section.
The bulk equation (\ref{bulk}), in combination with equations (\ref{MechanicalBoundCond})-(\ref{steady:eq3}) for the surface diffusion, or equations (\ref{mech_bound_cond_PT})-(\ref{steady:eq2}) for the phase transformation mechanism, describes the dynamics of the two models. 
In both cases they lead to a complicated free boundary problem.

Before starting to solve the full free boundary problem numerically, we first discuss qualitatively the existence of steady state solutions, by the use of scaling arguments. 
Here, the term steady state describes a non-equilibrium solution, for which the crack is moving with a constant velocity $\velo$, and in a co-moving frame of reference -- following the crack tip with the same steady state velocity $\velo$ -- the shape is constant in time. 
The following scaling arguments are fairly generic and can similarly be applied to both the SD and the PT model and predict the characteristic velocity and tip scale.

We will address the selection problem on different levels:
First, we use pure dimensional arguments for potential length- and velocity scales, and we show that only with the additional parameters stemming from viscoelasticity or mass density microscopic tip scales appear, which can select a tip radius.
This argument captures already the essential physical situation, and therefore the following, more extended discussion could be skipped for first reading.
There, we revisit the behavior with a more detailed analysis of the equations of motion;
the basic outcome of this more advanced investigation is that for the purely elastostatic case tip radius and velocity cannot be selected inpendently.
We then show how the inclusion of inertial or viscoelastic effects cures this problem.

The simplest assessment is to determine the possible nontrivial parameter combinations in the model in order to form a lengthscale.
For pure linear static elasticity, however, it is not possible to set a microscopic length scale out of the material parameters and the applied load, and therefore selection is not possible in this case.
The lack of such a lengthscale is the reason for the cusp singularity of the ATG instability and the impossibility of a steady state crack growth under these conditions.

The situation is different, when inertial effects are taken into account, because then the Rayleigh wave speed (the sounds wave speed at a free surface), $\velo_R\sim (E/\rho)^{1/2}$ appears as characteristic velocity scale in the problem.
Then we expect the crack velocity to be of this order, $\velo\sim\velo_R$, and we can form a microscopic lengthscale by the parameter combinations $(D_s/\velo_R)^3$ for SD and $D/\velo_R$ for PT, which we expect to be the tip scales in this case.
If instead of inertial effects viscous bulk damping becomes relevant, the characteristic velocity scale is $\velo_0\sim (D_s E^3/\eta^3)^{1/4}$ for SD and $\velo_0\sim (D E/\eta)^{1/2}$ for PT.
Consequently, we expect the characteristic tip scales $(D_s\eta/E)^{1/4}$ for SD and $(D\eta/E)^{1/2}$. 

For a more detailed analysis, we have to inspect the fields and the equations of motion.
We first return to the case without viscous or inertial effects.
We note that the stresses on the boundary of the crack
tip with finite radius $r_{0}$ scale as
\begin{align}
\sigma & \sim Kr_{0}^{-1/2},\label{sqrt_decay_stresses}
\end{align}
where $K$ is the stress intensity factor.
Since this analysis can be applied both to mode I and mode III, we do not indicate the loading conditions in the stress intensity factor.
Also, by boundary conditions the normal and shear stresses on the crack surfaces vanish (or are small, if the momentum transfer term is included for surface diffusion, provided that the crack speed is substantially smaller than the speed of sound).
Therefore, the only nontrivial stress component is the tangential stress, which depends on the shape of the entire crack.

Although the stress scaling (\ref{sqrt_decay_stresses}) appears to be natural in the framework of fracture mechanics, it is not trivial, and therefore we discuss it in more detail.
Since we intend to describe cracks with a finite tip radius, the stress field typically contains not only $r^{-1/2}$ terms but also faster diverging terms $r^{-3/2}, r^{-5/2}, \ldots$.
These terms cannot be present for sharp crack tips, since they would lead to a diverging elastic energy, but cannot be excluded for finite $r_0$, since then the divergency is cut off.
Therefore, the stress field typically consists of singular and regular parts
\begin{equation}
\sigma(r, \theta) \sim \frac{K}{(2\pi r)^{1/2}} \left( 1 + c_1 \frac{f^{(1)}}{r} +  c_2 \frac{f^{(2)}}{r^2} + \ldots \right) +\sigma_{reg}.
\end{equation}
in a polar representation.
For the sake of brevity we do not write the angular dependence $f^{(i)}$ explicitly.
The regular part of the stress $\sigma_{reg}$ contains only constant contribution (``$T$ stress'') and positive powers of $r$.
In the region $r_0\leq r \ll W$, i.e.~close to the (small) crack tip on the scale of the system size $W$, the ascending powers can however be ignored.
Also, the T stress scales as $K/W^{1/2}$, and therefore vanishes for large system sizes and given stress intensity factor $K$.
Consequently, $\sigma_{reg}=0$ for the further discussion.
The above representation is the heart for the multipole expansion method that will be introduced below in Section \ref{MEM}.
Notice that the higher order modes seem to violate the anticipated $r_0^{1/2}$ scaling of Eq.~(\ref{sqrt_decay_stresses}).
To understand this situation,
we inspect a crack with a different tip radius, which we obtain by a geometrical rescaling $r\to r/\alpha$, and with the same stress intensity factor.
We make the scaling ansatz for the stresses $\tilde{\sigma}$ for this rescaled geometry as $\tilde{\sigma}(r, \theta)=\beta \sigma(r/\alpha, \theta)$ in relation to the original problem.
At large distances from the tip, the main mode $\sigma\simeq K/(2\pi r)^{1/2}$ prevails, hence $\tilde{\sigma}(r, \theta) = \beta\sigma(r/\alpha, \theta)\simeq \beta\alpha^{1/2}K/(2\pi r)^{1/2}$.
On the other hand, both cracks look identical on large distances, where they become sharp straight cracks, and therefore $\beta=\alpha^{-1/2}$.
On the crack surface, where also the higher modes are relevant, the stress of the rescaled problem is therefore $\tilde{\sigma}(\tilde{r}_0, \theta) = \tilde{\sigma}(\alpha r_0,\theta)=\sigma(r_0, \theta)\alpha^{-1/2}$.
Hence, for a crack with a four times sharper tip, the surface stresses are two times higher, as stated in Eq.~(\ref{sqrt_decay_stresses}).

Obviously, the interface curvature in the tip region scales as $\kappa\sim1/r_{0}$.
Hence, as long as only {\em static linear elasticity} is taken into account, all contributions to the chemical potential scale as $\mu\sim r_{0}^{-1}$.
Consequently, a rescaling of the equations
of motion (\ref{steady:eq3}) or (\ref{steady:eq2}) is possible:
Formally, the equations of motion depend only on the dimensionless
combinations $\velo r_{0}^{3}/D_{s}$ for the SD mechanism
and $\velo r_{0}/D$ for the PT dynamics. All other
parameters combine to the dimensionless driving force $\Delta=K^{2}(1-\nu^{2})/2E\gamma$
(in case of pure mode I loading), where $\Delta=1$ corresponds to
the Griffith point. In other words, the radius $r_{0}$ and the steady
state velocity $\velo$ cannot be selected separately within the
framework of the pure static theory of elasticity.
Even if  a steady state solution exists, it would still degenerate to a one-parameter family of solutions with either fast cracks with small tip radii or slow cracks with large tip scales.
It turns out, however, that no steady state solutions exist, which is exactly the aforementioned cusp singularity of the ATG instability.


Additionally to this inspection of the behavior in the tip region, we can get more insights from the analysis of the crack shapes in the tail region, where the elastic stresses have decayed. 
To that end we assume that the crack is growing in the steady state regime in positive $x$-direction with a constant
velocity $\velo$. For the SD model, the shape equation
(\ref{steady:eq3}) can be integrated once, and in the co-moving frame
of reference we obtain
 \begin{align*}
\velo y & =-\frac{D_{s}}{\gamma\Omega}\frac{\partial\mu}{\partial s}.
\end{align*}
This is a complicated, non-linear third order equation with non-local
contributions arising from the elastic fields, since $\sigma_{ik}$
depends on the entire shape. In the tail region the shape equation is simplified to the third order
differential equation $D_{s}y'''=\velo y$ due to the decay of
the stress fields, which has two growing
and one decaying solution. Only the latter $y(x\to-\infty)=A\exp((\velo/D_{s})^{1/3}x)$
asymptotically describes the physically allowed shape. We switch to
a polar coordinate system $x=r(\theta)\cos\theta,y=r(\theta)\sin\theta$ for the tip region,
and focus on symmetrical solutions, $r(\theta)=r(-\theta)$. Since
the physical properties, curvature and stresses, do not depend on
the choice of the coordinate system but only on the crack shape, we
can arbitrarily chose $r(\theta=0)=r_{0}$, with the a priori unknown
tip radius $r_{0}=1/\kappa(0)$. Then from symmetry considerations
and the definition of the tip curvature, $\kappa=(r^{2}+2r'^{2}-rr'')/(r^{2}+r'^{2})^{3/2}$,
the natural conditions $r'(0)=r''(0)=0$ arise. Integration over the
upper interface $\theta>0$ requires the suppression of the two growing
exponentials at the tail, which imposes two boundary conditions. For
a given external loading, these two conditions can be fulfilled by
a proper selection of the tip radius $r_{0}$ and growth velocity
$\velo$. However, since the use of only linear static theory of
elasticity does not allow the independent selection of both the tip
radius $r_{0}$ and the steady state velocity $\velo$, the selection
will not suppress both growing exponentials at the same time, and
consequently a crack like solution does not exist \cite{BrenerSpatch2003}. 

For the PT model, a similar argument can be given
\cite{Spatchek2006,pilipenko2007}. In the tail region, where the
elastic stresses have decayed, the shape equation becomes simply $-\velo y'=Dy''$.
Its general solution, $y(x)=h+B\exp(-\velo x/D)$, contains the finite
crack opening $h$ and a growing exponential. Notice, that in contrast
to the surface diffusion process a finite opening $2h$ cannot be
excluded since we do not have to obey mass conservation here. Suppressing
the exponential and selecting a finite tail opening $h$ finally requires
again the independent selection of both the steady state propagation
velocity and the crack tip radius. Consequently, a steady state solution
for a growing crack in the framework of the static theory of elasticity
does not exist.

The situation changes if additional length scales enter into the description, and two natural aforementioned extensions for a description of crack growth are viscous bulk dissipation and dynamic elasticity. 
Nonlinear elastic or plastic effects are also conceivable, but are beyond the scope of the present article \cite{Langer072000, LoPomyalovProcaccia2008, BouchbinderLivne2008, LivneBouchbinder2008, Bouchbinder2009}.

Although the viscous stress defined in Eq.~(\ref{ViscousStresses})
introduces two new parameters, the timescales, which are introduced by them, should typically be of the same order.
By setting $\zeta=\nu$ we restrict to the case of only one additional
time scale $\tau=\eta/E$ to simplify the situation. Then, considering static elasticity and
viscous bulk dissipation, additionally the dimensionless ratio $\velo/\velo_{0}$ with $\velo_{0}=({D/\tau})^{1/2}$ for PT and $\velo_{0}=(D_{s}\tau^{-3})^{1/4}$
for SD appears
in the equations of motion, and therefore a rescaling is no longer
possible. Then this additional free parameter allows to independently
select both the tip scale $r_{0}$ and the steady state velocity $\velo$, so that the two growing exponentials can be suppressed.

To make this more explicit, we note that by virtue of Eqs.~(\ref{HookeLaw}) and (\ref{ViscousStresses}) for steady state growth $\sigma^{(vis)}=-\velo \tau\partial_x \sigma^{(el)}$.
Therefore, as consequence of the force balance condition (\ref{bulk}) we get a correction to the elastic stresses which depends on the dimensionless parameter $\velo \tau/r_0$.
For SD, in the interface evolution equation the two non-dimensional parameters $s_1=\velo r_0^3/D_s$ and $s_2=\velo \tau/r_0$ appear, which contain different combinations of the tip radius and the crack velocity, hence these two nonlinear eigenvalues can be selected independently to suppress the two growing exponential terms in the tail region.
Since $s_1$ and $s_2$ are of order unity for driving forces of order one, we therefore get $s_1s_2^3=\velo^4\tau^3/D_s\sim 1$, hence $\velo\sim (D_s\tau^{-3})^{1/4}$, which is the predicted velocity scale $\velo_0$.
Analogous arguments can be used for the PT mechanism.
Similarly, we obtain $v\sim v_R$ if inertial effects are relevant, and the microscopic tip scales are $r_0\sim(D_s/\velo_R)^3$ for SD and $r_0\sim D/\velo_R$ for PT \cite{BrenerSpatch2003,Spatchek2006,Spatschek2007}.


Altogether, we conclude that independent of the considered mass transport
mechanism, steady state growth of cracks is possible if apart from
static elasticity at least one additional effect is taken into account.
Furthermore, for both mechanisms a tip splitting is at least conceivable for high applied
tensions due to a secondary ATG instability: Since $\sigma\sim Kr_{0}^{-1/2}$
in the tip region and the local ATG length is $L_{G}\sim E\gamma/\sigma^{2}$,
an instability may occur, provided that the tip radius reaches by blunting the order
of the ATG length. 

The similarity of the scaling arguments to predict steady state growth for both SD and PT emphasizes the close connection between the two models.
From a physical point of view the SD model is probably more appealing, but also more difficult to solve numerically.
However, the preceding arguments suggests that many generic properties of the model should also be reflected in the simpler PT model.
In section \ref{resultssec} we will give a detailed comparison between the model behaviors.


\section{Numerical methods}
\label{methods}

The free boundary problem, which arises from the
coupling of non-local dynamic elasticity or viscoelasticity to the interface kinetics, is studied by the use of two complementary
methods, which are presented in this section. 

The first method is a sharp interface method, based on the expansion of
the elastic fields in a series of eigenfunctions of a straight mathematical
cut.
This multipole expansion method, designed to simulate efficiently steady state crack propagation, delivers precise
results in two limiting cases:
Slow crack propagation with viscoelastic effects and elastodynamic fracture
without bulk dissipation.
Situations where both effects play a role can only be treated in a perturbative manner.

As second method we use a fully
dynamic phase field model with a sharp interface limit corresponding
to our model equations.
In contrast to the multipole expansion method the phase field approach allows also to investigate transient behaviors and crack branching, and also enables to model both elastodynamic and viscoelastic effects in a uniform framework.
However, obtaining quantitative results
comparable with those by the multipole expansion method is computationally
very expensive.

\subsection{Multipole Expansion Method}
\label{MEM}


For solution of the steady state problem of crack propagation with the multipole expansion method we divide the problem into two parts:
First, the solution of the mechanical  problem for an arbitrary, but known crack shape and velocity $\velo$, and second the
evolution of the crack contour and adjustment of the velocity.
These two steps are iterated until a self-consistent steady state solution is found.


To simplify the appearance of viscosity in our equations, we assume $\zeta=\nu$
and thereby focus on the case of only one additional time scale $\tau=\eta/E$
due to viscosity.
With this simplification the dissipative stress tensor is related to the elastic stress tensor by $\sigma_{ik}^{(vis)}=\tau\dot{\sigma}_{ik}^{(el)}$.
We note that for mode III fracture such a simplifying parameter choice is not necessary, since there always only one timescale appears. 
For steady state growth, the time derivative in the co-moving frame is then replaced
by a spacial derivative with respect to the crack propagation direction
$x$, $\partial/\partial t=-\velo\partial/\partial x$.
Consequently, the steady state mechanical bulk equilibrium equations Eq.~(\ref{bulk}), containing both viscoelastic and inertial effects, become
\begin{equation}
\frac{\partial}{\partial x_{k}}\left(\sigma_{ik}^{(el)}-\tau\velo\frac{\partial}{\partial x}\sigma_{ik}^{(el)}\right) =\rho\velo^{2}\frac{\partial^{2}u_{i}}{\partial x^{2}}.
\label{steady_state_bulk}
\end{equation}

The basic idea for solving the elastic problem, is to write the elastic fields as an expansion in eigenfunctions of the differential operator corresponding to the equations of motion (\ref{steady_state_bulk}) for a straight moving cut.
Formally, the structure of the stress fields becomes
\begin{equation}
\sigma \sim \frac{K}{(2\pi r)^{1/2}} \left( 1 + c_1\frac{f^{(1)}}{r} + c_2\frac{f^{(2)}}{r^2} + \ldots \right),
\end{equation}
where the coefficients $c_i$ are the amplitudes of the eigenmodes $f^{(i)}$.
Then, the bulk equations are automatically fulfilled, and the problem is reduced to a linear one for finding proper expansion coefficients $c_i$ in order to satisfy the boundary conditions Eqs.~(\ref{MechanicalBoundCond}) or (\ref{mech_bound_cond_PT}). 
This reduction makes this method numerically very efficient.

However, to our best knowledge, there is no closed solution to the full problem, which means that the eigenfunctions for the visco-elastodynamic problem of a moving mathematical cut are not known.
Therefore, we focus here on two limiting cases of Eq.~(\ref{steady_state_bulk}):
First, the static limit of viscoelasticity, where $\velo\ll\velo_{R}$, and therefore the term on the right hand side is neglected, and second the elastodynamic limit where the viscous damping vanishes, i.e.~$\tau=0$.
Here, we mainly deal with mode I fracture, and therefore we illustrate the corresponding procedures to solve both the viscoelastic problem and the elastodynamic problem for these loading conditions.
The subsequent technical subsections for the determination of the eigenmodes for these two cases can be skipped for the first reading.
For mode III loading, which is mathematically simpler, similar approaches can be found;
in particular, the solution of the viscoelastic mode III problem is presented in \cite{SpatschekBrenerPilipenko2008}.

\subsubsection{Viscoelasticity}

First, we consider the limit of small crack velocities, i.e.~$\velo\ll\velo_{R}$.
In this limit the the term from inertia on the right hand side Eq.~(\ref{steady_state_bulk})
can be omitted. Therefore, the force equilibrium condition in the
static limit for the steady state situation reads: 
\begin{align}
\frac{\partial}{\partial x_{k}}\left(\sigma_{ik}^{(el)}-\tau\velo\frac{\partial}{\partial x}\sigma_{ik}^{(el)}\right) & =0.
\label{Force_Balance_Visco}
\end{align}
For the solution of this problem, we use Airy functions, which are well known in static elasticity.
Here, we generalize this approach to viscoelastic materials.


We begin with the treatment of a static elastic problem and  introduce for convenience a complex Airy function $\mathcal{U}(z)$, with $z=x+iy$.
The usual (real) Airy function is defined as its real part,
\begin{equation}
U(x,y) = \Re \left( \mathcal{U}(z) \right).
\end{equation}
The usual relations to obtain stresses are
\begin{align}
\sigma_{xx} & =\frac{\partial^{2}U}{\partial y^{2}};\;\sigma_{xy}=-\frac{\partial^{2}U}{\partial x\partial y};\;\sigma_{yy}=\frac{\partial^{2}U}{\partial x^{2}}.
\label{Airy_stresses}
\end{align}
Compatibility, i.e. the existence of a displacement field from which the elastic strains can be derived, is equivalent to
\begin{equation} 
\label{airy::eq2}
\Delta^2 U = 0.
\end{equation}
In most cases, the complex Airy function $\mathcal{U}$ is not analytic, and the reason is that its real part has to satisfy only the biharmonic equation and not the Laplace equation.
We therefore make the ansatz
\begin{equation} \label{airy::eq1}
\mathcal{U} = f + \overline{z} g,
\end{equation}
with $f(z)$ and $g(z)$ being analytic functions (apart e.g. from a branch cut for crack problems);
the bar denotes complex conjugation.
This means that with $f=f_1 + if_2$ and real functions $f_1(x,y), f_2(x,y)$ the Cauchy-Riemann equations hold:
\begin{equation}
\frac{\partial f_{1}}{\partial x} = \frac{\partial f_{2}}{\partial y},\qquad \frac{\partial f_{1}}{\partial y} = -\frac{\partial f_{2}}{\partial x}.
\end{equation}
With the above structure Eq.~(\ref{airy::eq1}) the biharmonic equation Eq.~(\ref{airy::eq2}) is automatically fulfilled.
Stresses can be expressed as
\begin{eqnarray}
\sigma_{xx} &=& \Re \left[ -f'' + 2g' - \overline{z} g'' \right] \label{airy::eq3}, \\
\sigma_{xy} &=& \Im \left[ f'' + \overline{z} g'' \right] \label{airy::eq4}, \\
\sigma_{yy} &=& \Re \left[ f'' + 2g' + \overline{z} g'' \label{airy::eq5}\right],
\end{eqnarray}
where the $'$ denotes the derivative with respect to $z$. 
The expression for displacements are
\begin{equation}
2\mu (u_x+iu_y) = (3-4\nu)g - z \overline{g'(z)} - \overline{f'(z)},
\end{equation}
with $\mu=E/2(1+\nu)$, and thus we get for the derivatives and strain componentents
\begin{eqnarray}
\epsilon_{xx} &=& \frac{1}{2\mu} \Re \left[ 2(1-2\nu)g' - \overline{z} g'' - f'' \right], \\
\epsilon_{yy} &=& \frac{1}{2\mu} \Re \left[ 2(1-2\nu)g' + \overline{z} g'' + f'' \right], \\
\frac{\partial u_{x}}{\partial y} &=& \frac{1}{2\mu} \Im \left[ -4(1-\nu)g' + \overline{z} g'' + f'' \right], \\
\frac{\partial u_{y}}{\partial x} &=& \frac{1}{2\mu} \Im \left[ 4(1-\nu)g' + \overline{z} g'' + f'' \right], \\
\epsilon_{xy} &=& \frac{1}{2\mu} \Im \left[ \overline{z} g'' + f'' \right].
\end{eqnarray}


As mentioned before, the total stress decomposes into an elastic and a viscoelastic contribution,
\begin{equation}
\sigma_{ik}^{(tot)} = \sigma_{ik}^{(el)} + \sigma_{ik}^{(vis)},
\end{equation}
where the latter is for steady state growth
\begin{equation}
\sigma_{ik}^{(vis)} = -v\tau \frac{\partial}{\partial x} \sigma_{ik}^{(el)},
\end{equation}
and consequently we have the force balance condition (\ref{Force_Balance_Visco}).
In principle, it is not required that elastic and viscous stress satisfy the force balance separately, but only Eq.~(\ref{Force_Balance_Visco}) must hold for the total stress.
Also, only the elastic fields needs to satisfy compatibility conditions.
However, as we will see, all fields fulfill these conditions even separately.

The ansatz is that both the elastic and the total stress field can be derived from Airy functions which satisfy the biharmonic equation.
In particular, we anticipate the following structure of the complex Airy functions
\begin{equation} \label{crack::eq3}
\mathcal{U}^{(tot)} = F+\overline{z}G,\qquad \mathcal{U}^{(el)} = f+\overline{z}g
\end{equation}
with analytic functions $f(z), g(z), F(z), G(z)$;
for the present crack problem these functions are not analytic everywhere, but have a branch cut along the negative real axis, see below.
As we have seen above, the real Airy functions $U^{(tot)}=\Re(\mathcal{U}^{(tot)})$, $U^{(el)}=\Re(\mathcal{U}^{(el)})$ then satisfy the biharmonic equation.
Stresses can be derived from Eqs.~(\ref{airy::eq3})-(\ref{airy::eq5}).

Provided that the following equation is fulfilled,
\begin{equation} \label{crack::eq4}
U^{(el)} -v\tau \frac{\partial}{\partial x} U^{(el)} = U^{(tot)},
\end{equation}
then the steady state force balance (\ref{Force_Balance_Visco}) is fulfilled and a valid elastic displacement field exists by construction.
We note that the complex Airy functions are not differentiable in the complex sense due to the appearance of the complex conjugate factor $\bar{z}$, and thus we cannot generalize $\partial_x \Re = \Re d/dz$.
However, with the above ansatz (\ref{crack::eq3}) the equation (\ref{crack::eq4}) is fulfilled if
\begin{equation} \label{crack::eq5}
f+\overline{z}g - v\tau (f'+g+\overline{z}g')=F+\overline{z}G
\end{equation}
holds.
Separating ``harmonic'' and ``biharmonic'' parts gives
\begin{eqnarray}
f-v\tau (f'+g) &=& F, \label{crack::eq6} \\
g-v\tau g' &=& G. \label{crack::eq7}
\end{eqnarray}
Again, if (\ref{crack::eq6}) and (\ref{crack::eq7}) are satisfied, then (\ref{crack::eq5}) is also valid.

We write the functions $F$ and $G$ now as series expansions in the set of eigenfunctions of a straight mode I cut,
\begin{eqnarray}
F &=& \sum_{m=-1}^{\infty} A_m z^{1/2-m}, \label{crack::eq11} \\
G &=& \sum_{m=0}^{\infty} B_m z^{1/2-m}. \label{crack::eq12}
\end{eqnarray}
Notice that the summations start from different values of $m$, because the function $G$ appears with an additional prefactor $\overline{z}$ in the complex Airy function.
The lowest value of $m$ corresponds to the main mode.
In order to have the correct mode I symmetry, the coefficients of expansion $A_m$ and $B_m$ are real.

The far field behavior is controlled by the term with the lowest value
of $m$, which is the main mode. 
On large distances $r$ from the
tip, the crack looks like a semi-infinite mathematical cut, and this is reflected by the proper $r^{-1/2}$ decay of the stresses in this purely elastic regime.
The prefactor of the main mode is therefore related to the stress intensity factor, thus
\begin{equation} \label{crack::eq9}
A_{-1} = \frac{K_I}{3\sqrt{2\pi}}.
\end{equation}
Also, we have the requirement that on the straight cut normal and shear stresses have to vanish, hence
\begin{equation} \label{crack::eq8}
B_0 = 3A_{-1}.
\end{equation}

We write now also the functions $f$ and $g$ as series
\begin{eqnarray}
f &=& \sum_{m=-1}^{\infty} \tilde{A}_m f_m \label{fsumeq} \\
g &=& \sum_{m=0}^{\infty} B_m g_m \label{gsumeq}
\end{eqnarray}
with analytical functions $f_m$ and $g_m$, which will be determined below.
We define
\begin{equation}
\tilde{A}_m = \left\{
\begin{array}{ll}
A_m & (A_m\neq 0), \\
B_m & (A_m=0).
\end{array}
\right.
\end{equation}
Notice that we assigned for convenience
\begin{equation}
B_{-1} = 0.
\end{equation}
Provided that
\begin{equation}
g_m -v\tau g_m' = z^{1/2-m} \label{crack::eq13}
\end{equation}
for $m=0, 1, 2, \ldots$, and either
\begin{equation}
f_m - v\tau \left( f_m' + \frac{B_m}{A_m} g_m \right) = z^{1/2-m} \label{crack::eq14}
\end{equation}
for $A_m\neq 0$ or
\begin{equation} 
f_m - v\tau f_m' = v\tau g_m, 
\end{equation}
for $A_m= 0$ hold for $m=-1, 0, 1, \ldots$, then Eqs.~(\ref{crack::eq6}) and (\ref{crack::eq7}) are satisfied.

Notice that the distinction between the regular case $A_m\neq 0$ and the singular case $A_m=0$ is relevant also for practical purposes:
For numerics we cut off the expansion of $F$ at $M-1$ and $G$ at $M$;
then $B_M\neq 0$, but $A_M=0$, so for the last mode we always encounter this situation.

Obviously, the equation for $f$ (\ref{crack::eq14}) can be solved as soon as the solution of the equation for $g$ (\ref{crack::eq13}) is known.


The homogeneous equation for $g$, $g_m^{(h)} - v\tau {g_m^{(h)}}'=0$ has the solution $g_m^{(h)}=D_m\exp(z/v\tau)$.
Variation of constant $D_m\to D_m(z)$ then gives the general solution of Eq.~(\ref{crack::eq13})
\begin{equation} \label{g::eq1}
g_m=D_m(z)\exp(z/v\tau)
\end{equation}
with
\begin{equation} \label{g::eq2}
D_m(z) = -\frac{1}{v\tau} \int \exp(-z/v\tau)z^{1/2-m} dz + const,
\end{equation}
where the integration constant has to be chosen such that exponential growth terms in (\ref{g::eq1}) are suppressed.
We obtain in particular for $m=-1$ with the proper integration constant
\begin{eqnarray}
D_{-1}(z) &=& \exp(-z/v\tau) \left[ z^{3/2} + \frac{3}{2} v\tau\, z^{1/2} \right] \nonumber \\
&& + \frac{3}{4} (v\tau)^{3/2} \sqrt{\pi} \erfc\sqrt{z/v\tau}
\end{eqnarray}
with the (complex) complementary error function $\erfc$.
Thus we obtain
\begin{eqnarray}
g_{-1}(z) &=& z^{3/2} + \frac{3}{2} v\tau\, z^{1/2} \nonumber \\
&& + \frac{3}{4} (v\tau)^{3/2} \sqrt{\pi} \exp(z/v\tau) \erfc\sqrt{z/v\tau}.
\end{eqnarray}
All higher modes can be obtained from the recursion relation
\begin{equation}
g_{m+1} = - \frac{1}{(\frac{1}{2}-m)v\tau} \left[ z^{1/2-m}-g_m \right], \label{gmp1eq}
\end{equation}
which follows from Eqs.~(\ref{g::eq1}) and (\ref{g::eq2}) and the proper choice of the integration constant.
In particular,
\begin{equation}
g_0 = z^{1/2} + \frac{1}{2} (v\tau)^{1/2} \sqrt{\pi} \exp(z/v\tau) \erfc\sqrt{z/v\tau}, \label{g0eq}
\end{equation}
which is the main mode term.



The equation for $f$, Eq.~(\ref{crack::eq14}), is treated in the same way, and we obtain
\begin{equation}
f_m = \left\{
\begin{array}{ll}
\displaystyle
g_m + \frac{B_m}{A_m} v\tau\, h_m \qquad & (A_m\neq 0) \\
\displaystyle
v\tau\, h_m \qquad & (A_m=0),
\end{array}
\right. \label{fmeq}
\end{equation}
with another analytical function $h_m(z)$.
It obeys the recursion relation
\begin{equation} \label{hmp1eq}
h_{m+1} =  \frac{1}{(\frac{1}{2}-m)v\tau} \left[ -g_m + h_m \right]
\end{equation}
and
\begin{eqnarray} \label{hm1eq}
h_{-1} &=& z^{3/2} + \frac{15}{4} v\tau\, z^{1/2} - \frac{3}{4}(v\tau)^{1/2}\sqrt{\pi} \times \nonumber \\
&& \times \left( z-\frac{5}{2} v\tau \right) \exp(z/v\tau) \erfc\sqrt{z/v\tau}.
\end{eqnarray}
In particular
\begin{eqnarray}
h_0 &=& \frac{3}{2} z^{1/2} - \frac{1}{2} (v\tau)^{-1/2}\sqrt{\pi} \times \nonumber \\
&& \times \left( z- \frac{3}{2}v\tau \right) \exp(z/v\tau)\erfc\sqrt{z/v\tau}.
\end{eqnarray}

To summarize, Eqs.~(\ref{gmp1eq})-(\ref{hm1eq}) provide
a series of eigenfunctions for the steady state equation of motion
Eq.~(\ref{Force_Balance_Visco}). From these eigenfunctions, via Eqs.~(\ref{fsumeq}) and (\ref{gsumeq}), together with (\ref{airy::eq3})-(\ref{airy::eq5})
the total stress field can be calculated as a function of the coefficients
of expansion $A_{0},A_{1},\ldots$ and $B_{1},B_{2},\ldots$. While
the main mode coefficients $A_{-1}$ and $B_{0}$ are given by Eqs.~(\ref{crack::eq9}) and (\ref{crack::eq8}),
the remaining coefficients $\{A_{i}\},\{B_{i}\}$ are now determined
in order to fulfill the conditions $\sigma_{nn}=\sigma_{ns}=0$ on
the actual crack contour ($n$ and $s$ are normal and tangential
directions respectively). The optimization of these expansion coefficients is equivalent
to finding the minimum of the function \begin{align}
R(\{A_{i}\},\{B_{i}\}) & =\int\left(\sigma_{nn}^{2}+\sigma_{ns}^{2}\right)\, ds\label{eq:Fracture-crack::eq10}\end{align}
 with respect to $\{A_{i}\},\{B_{i}\}$, where the integration is
performed along the crack contour.

\subsubsection{Elastodynamics}

Now we discuss the solution of the elastic boundary value problem
in the dynamic limit of vanishing viscous bulk dissipation, i.e.~$\tau=0$.
Therefore we briefly review the analysis given in \cite{pilipenko2007}.
Following Ref.~\cite{Rice1968,Freund:1998yq}, we introduce two real functions
$\phi(x,y,t)$ and $\psi(x,y,t)$ which are related to the displacements
$u_{i}$ as follows, \begin{align*}
u_{x}=\frac{\partial\phi}{\partial x}+\frac{\partial\psi}{\partial y}, & \quad u_{y}=\frac{\partial\phi}{\partial y}-\frac{\partial\psi}{\partial x}.\end{align*}
 Using the decompositions of the displacement field, the steady state
bulk equations (\ref{steady_state_bulk}) become homogeneous Laplace
equations \begin{align}
\frac{\partial^{2}\phi}{\partial x^{2}}+\frac{\partial^{2}\phi}{\partial y_{d}^{2}} & =0,\quad\frac{\partial^{2}\psi}{\partial x^{2}}+\frac{\partial^{2}\psi}{\partial y_{s}^{2}}=0,\label{steady_state_bulk_dynamics}\end{align}
where the coordinates perpendicular to the crack are rescaled by either
$y_{d}=\alpha_{d}y$ or $y_{s}=\alpha_{s}y$ for either the function
$\phi$ or the function $\psi$. Here, we have defined $\alpha_{d}^{2}=1-\velo^{2}/c_{d}^{2}$
and $\alpha_{s}^{2}=1-\velo^{2}/c_{s}^{2}$, where, $c_{d}=\sqrt{E(1-\nu)/\rho(1-2\nu)(1+\nu)}$
and $c_{s}=\sqrt{E/2\rho(1+\nu)}$ are the dilatational and shear
sound speeds respectively. Since we are looking for solutions obeying
the mode I symmetry, we propose the ansatz \begin{align}
\phi & =\sum_{n=0}^{\infty}A_{n}r_{d}^{3/2-n}\cos\left(\frac{3}{2}-n\right)\theta_{d};\label{dyn_series_phi}\\
\psi & =-\sum_{n=0}^{\infty}B_{n}r_{s}^{3/2-n}\sin\left(\frac{3}{2}-n\right)\theta_{s},\label{dyn_series_psi}\end{align}
 in rescaled polar coordinates, which are related to the co-moving
Cartesian coordinates via $x=r_{d}\cos\theta_{d}=r_{s}\cos\theta_{s}$,
$y_{d}=r_{d}\sin\theta_{d}$ and $y_{s}=r_{s}\sin\theta_{s}$. For
a crack with a sharp tip, only the mode with $n=0$ is allowed, which
corresponds to the usual $\sigma\sim r^{-1/2}$ behavior. For this
mode, the boundary conditions on the straight cut and the matching
to the far field behavior demand \begin{align}
A_{0} & =\frac{8(1+\nu)(1+\alpha_{s}^{2})}{\sqrt{2\pi}3E(4\alpha_{s}\alpha_{d}-(1+\alpha_{s}^{2})^{2})}K_{I}^{dyn},\label{coefA}\\
B_{0} & =\frac{2\alpha_{d}}{1+\alpha_{s}^{2}}A_{0},\label{coefB}\end{align}
 where $K_{I}^{dyn}$ is the dynamical mode I stress intensity factor
\cite{Freund:1998yq}, related to the static stress intensity factor as
\begin{align}
K_{I}^{dyn} & =K_{I}^{stat}\left((1-\nu)\frac{4\alpha_{s}\alpha_{d}-(1+\alpha_{s}^{2})^{2}}{\alpha_{d}(1-\alpha_{s}^{2})}\right)^{1/2}.\end{align}
 Each eigenmode of Eqs.~(\ref{dyn_series_phi}) and (\ref{dyn_series_psi})
satisfies the elastodynamic bulk equation (\ref{steady_state_bulk_dynamics}).
The coefficients $A_{0}$ and $B_{0}$ are determined by Eqs.~(\ref{coefA})
and (\ref{coefB}) for the correct far-field behavior, whereas all
other modes decay faster. Consequently, we obtain the formal stress
field expansion, \begin{align}
\sigma_{ik} & =\frac{K_{I}^{dyn}}{(2\pi r)^{1/2}}\left(f_{ik}^{(0)}+\sum_{n=1}^{N=\infty}\frac{A_{n}f_{ik,d}^{(n)}+B_{n}f_{ik,s}^{(n)}}{r^{n}}\right),\label{sigma:exp}\end{align}
 where $f_{ik,d}^{(n)}(\theta,\velo)$ and $f_{ik,s}^{(n)}(\theta,\velo)$
are the universal angular distributions for the dilatational and shear
contributions which also depend on the propagation velocity.
In analogy to the procedure above, the unknown coefficients of the series expansion are determined by minimization of the residuum
\begin{equation}
R(\{A_i\}, \{B_i\}) =\int\left(\sigma_{ni}+\rho \velo_n \dot{u}_i\right)^2\, ds
\end{equation}
with respect to the coefficients $A_i$ and $B_i$, for a given crack contour and steady state velocity;
the integration domain is the crack contour.
Notice that this residuum is used for SD, whereas for PT it is the same as in Eq.~(\ref{eq:Fracture-crack::eq10}).

\subsubsection{Approximative viscoelastodynamic model}
\label{toy_model}

To our best knowledge, there is no exact solution of the full problem given by Eq.~(\ref{steady_state_bulk}), which contains both dynamical and viscous effects.
We therefore suggest an approximative model, which captures essential physical aspects and gives exact results both in the limit of vanishing viscous damping and treats viscous damping in the sense of a rigorous perturbation theory for low velocities.

To motivate the model let us first consider the case of static elasticity, where inertial effects can be neglected.
Thus, we solve the elastic problem consisting of three ingredients, i.e.~bulk equilibrium $\partial\sigma^{(el, 0)}_{ij}/\partial x_j=0$, stress free boundaries at the crack surfaces, $\sigma_{in}^{(el, 0)}=0$ and $\sigma^{(el, 0)}\sim K_{I} r^{-1/2}$ at large distances, and additionally compatibility, which means that the strain tensor, which is related to the stress via Hooke's law, can be derived from a displacement field.
We use here the superscript $(el, 0)$ to indicate that we are dealing here with a purely elastic field, which is later used as zeroth order for a perturbative treatment.
By these requirements the solution for $\sigma_{ij}^{(el, 0)}$ is formally uniquely defined (apart from translational and rotational degrees of freedom).

On the other hand, for the viscoelastic steady state problem the total stress consists additively of the elastic and viscous stress, $\sigma_{ij}^{(tot)}=\sigma_{ij}^{(el)} + \sigma_{ij}^{(vis)}$, where the viscous stress is related to the elastic stress by $\sigma_{ij}^{(vis)}=-\velo\tau\partial_{x}\sigma_{ij}^{(el)}$ (we assume here again that we have only one viscoelastic timescale and that the crack moves in positive $x$ direction).
Now the {\em total} stress has to fulfill mechanical equilibrium, $\partial\sigma_{ij}^{(tot)}/\partial x_j=0$, the {\em total} normal and shear stresses have to vanish on the crack surfaces, $\sigma_{in}^{(tot)}=0$ and far away (where the behavior is anyway purely elastic) we have the same asymptotic behavior $\sigma_{ij}^{(tot)}\sim K_{I} r^{-1/2}$.
Also, as we have seen in the section on the multipole expansion method, also the {\em total} stress field satisfies compatibility.
Therefore, by uniqueness of the solution, the solution for the {\em total} stress is here exactly the same as before for the purely elastostatic problem, $\sigma_{ij}^{(tot)}=\sigma_{ij}^{(el, 0)}$.

Now we use the solution of the elastic problem and introduce viscosity perturbatively, where we use $\tau$ as small expansion parameter.
The zeroth order solution $\sigma_{ij}^{(el,0)}$ generates a viscous stress to first order in $\tau$, i.e. $\sigma_{ij}^{(vis,1)}=-\velo\tau\partial_{x}\sigma_{ij}^{(el, 0)}$.
However, up to first order order the sum of these two terms, $\sigma_{ij}^{(el, 0)}+\sigma_{ij}^{(vis,1)}$ does not yet satisfy boundary conditions on the crack surfaces (but they do satisfy the bulk force balance conditions), and therefore another elastic correction term must appear to first order, $\sigma_{ij}^{(el,1)}$.
Hence, up to first oder the total stress is $\sigma_{ij}^{(tot)} = \sigma_{ij}^{(el,0)}+\sigma_{ij}^{(el,1)}+ \sigma_{ij}^{(vis,1)} +\mathcal{O}(\tau^2)$.
On the other hand, by the aforementioned uniqueness of the solution, $\sigma_{ij}^{(tot)}=\sigma_{ij}^{(el,0)}$ to all orders.
Consequently, we obtain $\sigma_{ij}^{(el,1)}=-\sigma_{ij}^{(vis,1)}=+\velo\tau\partial_{x}\sigma_{ij}^{(el,0)}$.

For the equations of motion we need an expression for the chemical potential, which depends on the elastic part of the stress only.
Hence we get a first order correction to the chemical potential
\begin{equation}
\Delta \mu^{(1)} = \Omega \velo\tau\epsilon_{ij}^{(0)}\frac{\partial\sigma_{ij}^{(el, 0)}}{\partial x}.\label{toy_model_idea}
\end{equation}
Here we have made use of the property $\sigma_{ij}^{(el, 0)}\epsilon_{ij}^{(1)}=\sigma_{ik}^{(el, 1)}\epsilon_{ik}^{(0)}$, which follows from Hooke's law.
Notice that there is no need to decorate the strain with a superscript $(el)$, since strain is by definition an elastic property (the viscous stress is related to the strain {\em rate}).
This description is a rigorous perturbative treatment of the viscoelastic theory in the quasistatic limit, $\velo\ll\velo_R$.

However, this concept cannot strictly be extended to the case of dynamical elasticity, since there on the right hand side of the Newtonian equation the acceleration term $\rho\ddot{u}_i$ contains only {\em elastic} displacements (so even in the force balance equation for the {\em total} stress the right hand side contains the {\em elastic} accelerations), and therefore the purely {\em elastodynamic} and the {\em total viscoelastodynamic} stresses do not obey the same equations.

Instead, we use the above recipe (\ref{toy_model_idea}) to incorporate viscous damping in the chemical potential also with inertial effects, and consider this as an approximative viscoelastodynamic model;
of course, this model is not rigorously derived, but captures essential aspects of the physics and is still exact for $\tau=0$ and becomes a rigorous perturbation theory for $\velo\ll \velo_R$.
Thus, in the framework of this model the chemical potential for the PT model is \begin{align}
\mu & =\Omega\left(\frac{1}{2}\sigma_{ik}^{(0)}\epsilon_{ik}^{(0)}+\velo\tau\epsilon_{ik}^{(0)}\frac{\partial\sigma_{ik}^{(0)}}{\partial x}-\gamma\kappa\right),\end{align}
where the stresses $\sigma_{ik}^{(0)}$ and strains $\epsilon_{ik}^{(0)}$
are calculated from the elastodynamic eigenfunctions (\ref{dyn_series_phi})--(\ref{dyn_series_psi}).
Notice, that for the SD mechanism, we also have to account for the kinetic energy density as in Eq.~(\ref{steady:eq1}). 


\subsubsection{Steady state crack growth}

Once we can solve the mechanical problem for arbitrary shape, we can
solve the free boundary problem for the steady state crack propagation.
The latter is described, depending on the mechanism of propagation,
by the set of Eqs.~(\ref{steady:eq1})--(\ref{steady:eq2}) in case
of the PT model or by Eqs.~(\ref{steady:eq1})--(\ref{steady:eq3})
in case of SD. The strategy for solving the problem
is as follows: For a given guessed initial crack shape and velocity,
we determine the unknown coefficients $A_{n}$ and $B_{n}$ from the
boundary conditions. Afterwards, we calculate the chemical potential
and the normal velocity at each point of the interface. The new shape
is obtained by advancing the crack according to the local interface
velocities. This procedure is repeated until the steady state is reached,
which means that the shape of the crack in the co-moving frame of
reference remains unchanged \cite{SaiGolHMK1988}. This ``quasi-dynamical'' approach provides
a natural way to solve the problem, as it follows the physical
configurations to reach the steady state. Then the following relation
between the local normal velocity and the steady state velocity $\velo$
holds
\begin{align}
\velo_{n}-\velo n_{x} & =0,
\label{ss}
\end{align}
 where $n_{x}$ is the $x$ component of the normal vector pointing into the solid phase. This is a purely geometrical relation and therefore
it is independent on the mechanism of crack propagation, i.e.~it
is valid for the SD model as well as for the PT model. This equation
gives us an alternative approach to the ``quasi-dynamical approach''.
Namely, we directly solve the nonlinear equation (\ref{ss}) as a
functional of the crack shape and the tip velocity $\velo$ by Newton's
method complemented by Powell's hybrid method \cite{Powell1970,DennisShnabelShnabel1983}
and we refer to this as the ``steady state approach''. We stress
here that the ``steady state approach'' is  preferable especially
in case of the SD model, where we thus can avoid to solve the fourth
order differential equation (\ref{steady:eq3}). 

Finally, we define the dimensionless driving force \begin{align}
\Delta & =\Delta_{I}+\Delta_{III}=\frac{1-\nu^{2}}{2E\gamma}K_{I}^{2}+\frac{1+\nu}{2E\gamma}K_{III}^{2},\label{dimless-driving-force}\end{align}
 where we also include the possibility of mixed-mode loading.
Here, $\Delta=1$ corresponds to the Griffith point, and the energetics
necessarily require $\Delta>1$ for crack growth.

\subsection{Phase Field modeling of fracture}

During the past years, phase field modeling has emerged as a promising
approach to model crack propagation by continuum methods (see \cite{SpatschBrenerKarma012010} for a recent review). This
method is especially advantageous due to its high versality to study
quite complicated crack patterns as well as multi crack situations \cite{SpatschGugenbeBrener102009}.
Nowadays, phase field models capture many known features of cracks \cite{KarmaKesslerLevine2001,HenLev2004,AraKalVin2000,EasSetRau2002};
However, a significant attribute of most of these descriptions is that the scale
of the growing patterns is always set by the phase field interface
width, which is a purely numerical parameter and not directly connected
to physical properties; therefore these models do not possess a valid
sharp interface limit. Alternative descriptions, which are intended
to investigate the influence of elastic stresses on the morphological
deformation of surfaces due to phase transition processes, are based
on macroscopic equations of motion. But they suffer from inherent
finite time singularities which do not allow steady state crack growth
unless the tip radius is again limited by the phase field interface
width \cite{Kassner2001,Spatschek2007}.

Since the phase field method was originally developed to mainly simulate
the dynamics of solidification processes, it is of course more natural
to formulate a phase field model for fracture using the PT
mechanism Eq.~(\ref{steady:eq2}). However, we mention here that
it is also possible to formulate phase field models for crack propagation
by surface diffusion \cite{RaeVoi07,gugenbergerSpatKassner2008,KassnerSpatschGugenbe042010},
and for example the initial stage of the ATG-instability has already
been reproduced using such kind of phase field models. Nevertheless,
for the current purpose, we restrict the discussion to phase field
modeling for crack propagation using non-conserved order parameter
dynamics, see Eq.~(\ref{steady:eq2}).

For the formulation of the present phase field model, we start with
the introduction of a continuous phase field $\phi$, which will discriminate
between the different material states. We define $\phi=1$ for the viscoelastic
medium, and $\phi=0$ for the ``broken phase''. 
This region does not support elastic stresses (the material is broken), but still it has the same density as the surrounding matrix.
Therefore, we use the notation of a ``dense gas phase'' to underline that we do not have vacuum inside the crack.

We start from
a free energy functional, similar to \cite{Kassner2001} 
\begin{align}
F[\phi,u_{i}] & =\int\limits _{V}\left(f_{s}+f_{dw}+f_{el}\right)dV,
\label{eq:CrackPropagation-free-energy-functional} 
\end{align}
where $f_{s}(\nabla\phi)=3\gamma\xi(\nabla\phi)^{2}/2$ is the gradient
energy density and $f_{dw}(\phi)=6\gamma\phi^{2}(1-\phi)^{2}/\xi$
is the double well potential, guaranteeing that the free energy functional
has two local minima at $\phi=0$ and $\phi=1$ corresponding to the
two distinct phases of the system. The form of the double well potential
and the gradient energy density are chosen such that the phase field
parameter $\xi$ defines the interface width and the parameter $\gamma$
corresponds to the interface energy of the sharp interface description
\cite{gugenbergerSpatKassner2008}. Finally the elastic energy density
contribution is 
\begin{align}
f_{el} & =\frac{h(\phi)E}{2(1+\nu)}\left(\frac{\nu}{1-2\nu}\epsilon_{ii}^{2}+\epsilon_{ik}^{2}\right),\label{PhaseField:free-energy}
\end{align}
 where $h(\phi)=\phi^{2}(3-2\phi)$ interpolates the elastic modulus
between zero for the dense gas phase and one for the viscoelastic
medium. It is the simplest polynomial satisfying the necessary interpolation
conditions $h(0)=0$ and $h(1)=1$ and having a vanishing slope at
$\phi=0$ and $\phi=1$, in order not to shift the bulk states. Here,
for the sake of brevity we consider the Poisson ratio to be phase
independent. 

The evolution of the elastic fields is determined by the principle
of momentum conservation according to Eq.~(\ref{bulk}), \begin{align}
\rho\ddot{u}_{i} & =\frac{\partial}{\partial x_{i}}(\sigma_{ik}^{(el)}+\sigma_{ik}^{(vis)}),\label{phase:eq1}\end{align}
 where the elastic stresses are defined as the derivative of the elastic
free energy density with respect to the strains, i.e.~$\sigma_{ik}^{(el)}=\partial f_{el}/\partial\epsilon_{ik}$.
In order to have vanishing viscous damping inside the crack, we use define the viscosity parameter $\eta$ to be phase field dependent, i.e. $\eta\to\eta h(\phi)$ in Eq.~(\ref{ViscousStresses}), while the parameter $\zeta$ remains phase independent.

The phase field dynamics is related to the functional derivative of
the free energy with respect to the phase field variable, \begin{align}
\frac{\partial\phi}{\partial t} & =-\frac{D}{3\gamma\xi}\left(\frac{\delta F}{\delta\phi}\right)_{u_{i}=const.},\label{phase:eq2}\end{align}
 where $D$ corresponds to the above mentioned kinetic coefficient
of the phase transformation model. Here it is assumed that the viscous dissipation
does not affect the phase field dynamics. According to our sharp interface
model of crack propagation, we consider viscosity to be a bulk property,
which does not affect the phase change behavior directly. 
We ignore local heating effects through bulk or interfacial dissipation, assuming that the heat diffusion is sufficiently fast.

Using the phase field method, we investigate crack growth in a strip
geometry with fixed displacements at the upper and lower grip.
In contrast, the
multipole expansion technique \cite{SpatschekBrenerPilipenko2008,pilipenko2007}
is designed to model a perfect separation of the crack tip scale  
 to the strip width $W$, i.e. $W\gg D/\velo_{R}$ or $W\gg D/\velo_{0}$, respectively. In most real cases, crack
tips are very tiny, and therefore it is theoretically desirable to
describe this limit. For the phase field method, however, a finite
strip width $W$ is necessary, and a good separation of the scales
therefore requires time-consuming large-scale calculations. We shift
the system such that the tip remains in the horizontal center. This
allows to study the propagation for long times until the crack reaches
a steady state situation. Apart from this finite size restriction,
introduced the interface width $\xi$ as a numerical parameter,
and the phase field method delivers quantitative results only in the
limit that all physical scales are much larger than this lengthscale.
The latter has to be noticeably larger than the numerical lattice
parameter $\Delta x$, but the results show that the choice $\xi=5\Delta x$
is sufficient. Therefore, to obtain quantitative agreement with the
results from the multipole expansion method, we have to satisfy the
hierarchy relation \begin{align}
\xi & \ll\frac{D}{\velo_{R}}\ll W\;\;\;\mbox{ or }\;\;\; \xi \ll \frac{D}{v_0} \ll W,\label{ScaleSeparation}\end{align}
 which is numerically very hard to achieve. 
 
We developed a parallel version of the phase field code which is running on up to 2048 processors,
with system sizes up to $8192\times4096\cdot(\Delta x)^{2}$ ($\Delta x$ is the lattice unit). 
However, for qualitative results we typically use $W\velo_{R}/D=86$ and $D/\velo_{R}\xi=1.9$,
where the total size of the system in grid points is $2048\times800$.
All computations are performed on the supercomputer JUGENE operated at the Research Center J{\"u}lich.

The dimensionless driving force $\Delta$ decomposes into mode I and
III contributions, and according to Eq.~(\ref{dimless-driving-force}),
it is defined for the strip geometry as \begin{align}
\Delta & =\Delta_{I}+\Delta_{III}=\frac{E}{2\gamma W}\left(\frac{\delta_{I}^{2}}{2(1-\nu^{2})}+\frac{\delta_{III}^{2}}{1+\nu}\right).\label{dimless-driving-force-phase-field}\end{align}
 Here, $\delta_{I}$ is the above mentioned fixed displacement by
which the strip is elongated vertically, whereas $\delta_{III}$ is
a fixed displacement by which the strip is sheared in the $z$ direction.
The value $\Delta=1$ corresponds to the Griffith point. 

\section{Results}
\label{resultssec}

In the following section we will give a comprehensive overview on the different results.
As has been mentioned above, we consider two different material transport mechanisms, surface diffusion (SD) and a phase transformation process (PT).
From a theoretical point of view, two different physical mechanisms, i.e.~viscous dissipation and inertial effects are important to provide selection mechanisms for the steady state velocity and the crack tip structure.
These two cases can be considered as limiting situations for slow and fast cracks, and here a quantitative treatment not only with phase field but also sharp interface methods is feasible (multipole expansion method).
The cross-over behavior, where both effects are relevant, is modeled using perturbation techniques for the multipole expansion method and fully dynamical phase field simulations.
Furthermore, apart from steady state growth also branching instabilities can occur, which will also be discussed.
Finally, we consider different loading modes, and we will start the discussion of the results for pure mode I fracture, before in the following section mixed mode situations with a combination of mode I and mode III loading are investigated.

Apparently, the different physical situations and numerical methods lead to a certain complexity of the results.
A concise summary of the results is therefore given in Table \ref{table1}.
\begin{table*}
\begin{tabular}{>{\raggedright}p{1.5cm}|>{\centering}p{7.2cm}|>{\centering}p{7.2cm}}
   & {\bf Surface Diffusion (SD)} & {\bf Phase transformation (PT)}  \tabularnewline
&&\tabularnewline
 \hline
 &&\tabularnewline
&
\begin{minipage}{7cm}
{\bf Viscoelastic limit: $\chi=\infty$}
\begin{itemize}
\item Creep branch: $1<\Delta<2.6$
\item Regular growth for $\Delta>2.6$
\begin{itemize}
\item Velocity grows monotonically with driving force
\item Velocity scale $\velo_0 \sim (D_s \tau^{-3})^{1/4}$
\item Tip scale $h_0\sim (D_s\tau)^{1/4}$
\item No branching
\end{itemize}
\end{itemize}
\end{minipage}  & 
\begin{minipage}{7cm}
{\bf Viscoelastic limit: $\chi=\infty$}
\begin{itemize}
\item Creep branch: $1<\Delta<2.6$
\item Regular growth for $\Delta>2.6$
\begin{itemize}
\item Velocity grows monotonically with driving force
\item Velocity scale $\velo_0 \sim (D/\tau)^{1/2}$
\item Tip scale $h_0\sim (D \tau)^{1/2}$
\item No branching
\end{itemize}
\end{itemize}
\end{minipage}  \tabularnewline
&&\tabularnewline
\cline{2-3}
&&\tabularnewline
{\bf Mode I
} &
\begin{minipage}{7cm}
{\bf Inertial limit $\chi=0$}
\begin{itemize}
\item No physically relevant solution
\end{itemize}
\end{minipage} &
\begin{minipage}{7cm}
{\bf Inertial limit $\chi=0$}
\begin{itemize}
\item No self-consistently selected tip radius in the range $1<\Delta<1.14$
\item Steady state solution for $\Delta>1.14$
\begin{itemize}
\item Velocity decaying function of $\Delta$
\item Velocity scale $\velo_R \sim (E/\rho)^{1/2}$
\item Tip scale $h_0\sim (D/\velo_R)^{1/2}$
\end{itemize}
\item Crack branching for $\Delta>1.8$
\end{itemize}
\end{minipage}  \tabularnewline
&&\tabularnewline
\cline{2-3}
&&\tabularnewline
%
&
\begin{minipage}{7cm}
{\bf Viscoelastodynamic regime $0<\chi<\infty$}
\begin{itemize}
\item Creep branch for low driving forces
\item Velocity first increases with $\Delta$, then decrease
\item No steady state solutions beyond a critical driving force, afterwards branching expected
\item Wide range of $\Delta$ for steady state solutions
\item Higher viscosity leads to lower crack speeds
\item Range of steady state solutions larger for higher viscous damping
\end{itemize}
\end{minipage}  &

\begin{minipage}{7cm}
{\bf Viscoelastodynamic regime $0<\chi<\infty$}
\begin{itemize}
\item Creep branch for low driving forces
\item Velocity first increases with $\Delta$, then decrease
\item No steady state solutions beyond a critical driving force, afterwards branching expected
\item Small range of $\Delta$ for steady state solutions
\item Higher viscosity leads to lower crack speeds
\item Range of steady state solutions larger for higher viscous damping
\end{itemize}
\end{minipage} \tabularnewline
&&\tabularnewline
\hline
&&\tabularnewline

{\bf Mode III
} &

\begin{minipage}{7cm}
{\bf Viscoelastic limit $\chi=\infty$}
\begin{itemize}
\item Strong blunting below $\Delta=1.1$ (ductile-to-brittle transition)
\item Steady state regime: $1.1<\Delta<3.8$
\begin{itemize}
\item Velocity grows monotonically for $1.1<\Delta<3.5$
\item Velocity decays for $3.5<\Delta<3.8$
\item Velocity scale $\velo_0 \sim (D_s \tau^{-3})^{1/4}$
\item Tip scale $h_0\sim (D_s\tau)^{1/4}$
\end{itemize}
\item No steady state solutions for $\Delta>3.8$, where branching is expected
\end{itemize}
\end{minipage} &

\begin{minipage}{7cm}
{\bf Viscoelastic limit $\chi=\infty$}
\begin{itemize}
\item Logarithmic opening of the crack
\end{itemize}
\end{minipage} \tabularnewline
&&\tabularnewline
\hline
&&\tabularnewline
{\bf Mode I + III
} &

\begin{minipage}{7cm}
{\bf Viscoelastic limit $\chi=\infty$}
\begin{itemize}
\item Higher mode I contribution lead to shift of onset of branching towards higher $\Delta$
\item Creep branch with a low velocity plateau
\item For low $\Delta$ faster growth for higher mode III contribution may enable development of crack front instability
\end{itemize}
\vspace{0.5cm}
\end{minipage} &

\begin{minipage}{7cm}
{\bf Viscoelastic limit $\chi=\infty$}
\begin{itemize}
\item  Logarithmic opening of the crack.
\end{itemize}
\end{minipage}
\tabularnewline
\end{tabular}
\caption{Brief summarizing comparison of the different growth modes.}
\label{table1}
\end{table*}

\subsection{Opening Mode fracture}
\label{mode1} 

In this section we discuss exclusively mode I fracture in the different variants of the model.
As discussed before, selection is of central interest for this pattern formation aspect of fracture, and two principal mechanisms have been introduced before, which the selection through viscoelastic bulk damping and the inertia limitation of the crack speed.
In all following calculations we use $\zeta=\nu=1/3$.
Then, $\tau=\eta/E$ is the only remaining viscous time scale, and we define the dimensionless viscosity strength $\chi=\velo_{R}^{2}/\velo_{0}^{2}$, where $\velo_{R}$ is the Rayleigh speed and $\velo_{0}=(D_{s}\tau^{-3})^{1/4}$ for the SD model, while $\velo_{0}=({D/\tau})^{1/2}$ is used for the PT model.

First, we deal with slow crack propagation with a steady state velocity much smaller then the Rayleigh speed, i.e.~$\velo\ll\velo_{R}$.
In this case dynamic effects are negligible, and the application of
static elasticity is legitimate, $\chi=\infty$.
Next, we discuss the limit of fast crack propagation with vanishing viscous dissipation, where the steady state velocity $\velo$ and the finite tip radius $r_{0}$ are selected by dynamic effects only, $\chi=0$.
A more general situation, which contains both effects, will be discussed in the framework of a perturbation analysis using the multipole expansion method and fully dynamical phase field simulations, as well as non-steady state crack growth with crack branching.

The different kinetic mechanisms PT and SD lead to very similar results in general, apart from the fact that surface diffusion implies material conservation, and therefore the crack shapes differ (compare Figs.~\ref{GeometrySD} and \ref{GeometryPT}).
However, an important difference is that for surface diffusion steady state physically relevant solutions do not exist without viscoelastic damping.



\subsubsection{Slow cracks}
\label{slow_cracks}

In this regime it is assumed that the sound speed is much larger than the crack velocity, and therefore inertial effects can be neglected.

\begin{figure}
\begin{center}
\includegraphics[width=1\linewidth]{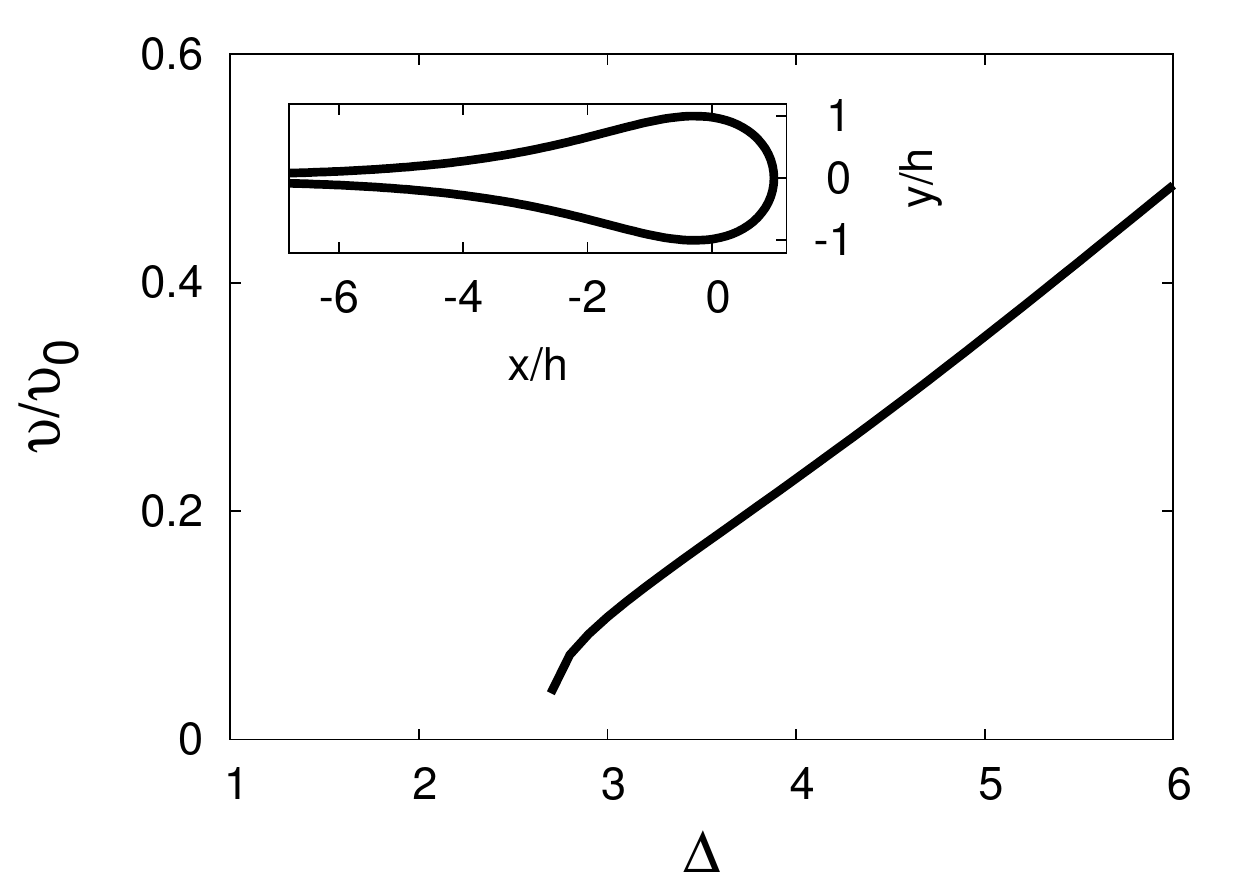} 
\caption{Steady state tip velocity $\velo/\velo_{0}$ for a mode I crack as
a function of the driving force $\Delta$ in case of the SD model.
The results are obtained with the multipole expansion method in the viscous limit.
The inset shows the corresponding crack shape for $\Delta=10.0$.
Both directions $x$ and $y$ are scaled with the half maximum height
$h$ of the crack. 
}
\label{mode1_visco_SD} 
\end{center}
\end{figure}

\begin{figure}
\begin{center}
\includegraphics[width=1\linewidth]{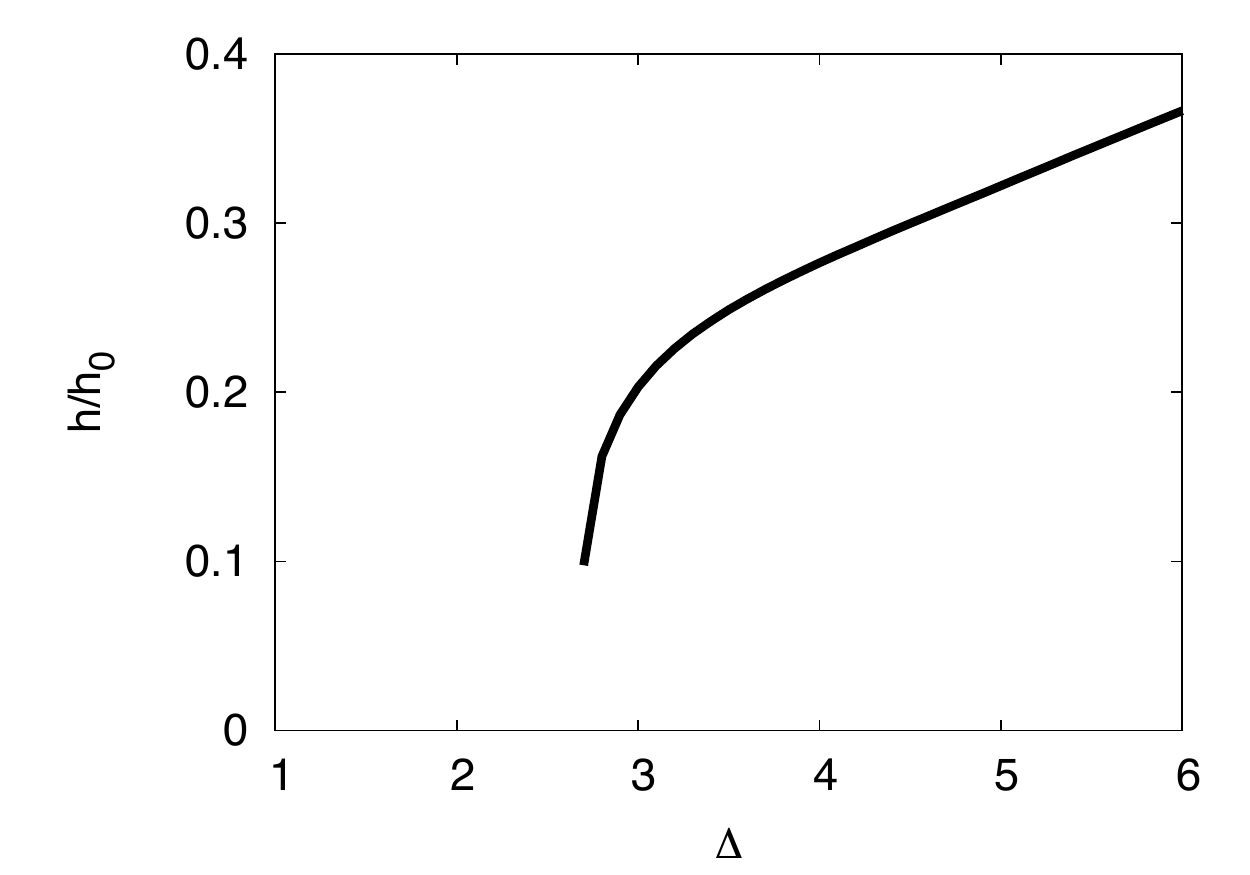}
\caption{The crack height, as defined in the inset of Fig.~\ref{mode1_visco_SD}, for the SD model in the viscoelastic limit. The characteristic lengthscale is defined as $h_0=(D_s\tau)^{1/4}$.
Below the value $\Delta=2.6$ the crack tip size becomes small, and simultaneously the crack velocity drops to very small values (``creep branch'');
then almost all dissipation stems from viscous bulk damping.
}
\label{ViscoSDheight}
\end{center}
\end{figure}

We start with reviewing the results for surface diffusion, as presented in \cite{SpatschekBrenerPilipenko2008}.
As for all surface diffusion models, only multipole expansion technique results are available, since the modeling of surface diffusion with phase field methods is more cumbersome \cite{RaeVoi07,gugenbergerSpatKassner2008,KassnerSpatschGugenbe042010}.
The numerical results for mode I fracture, as obtained from the simulations, are shown in Fig.~\ref{mode1_visco_SD} and \ref{ViscoSDheight}.
The inset of Fig.~\ref{mode1_visco_SD} shows a typical steady state crack shape, which has the characteristic features of a finite tip radius and vanishing surface separation in the tail region, which results from the material conservation condition, as discussed before.
We point out that the crack is shown in the Lagrangian reference frame, i.e.~without elastic displacements, as it would appear if suddenly the mode I loading was removed.
The maximum vertical diameter of the crack defines here the tip scale $2h$.
The velocity plot Fig.~\ref{mode1_visco_SD} shows only finite velocities above $\Delta\approx 2.6$, and from there on a strictly monotonic increase of the steady state velocity as function of the driving force.
We did not find any indications that this solution branch terminates at higher driving forces.
Notice that the crack speed is set by the characteristic viscoelastic velocity $v_0=(D_s\tau^{-3})^{1/4}$.
Reducing the driving force coming from high values, the crack velocity rapidly drops to very small values at $\Delta\approx 2.6$, and below this value the crack growth velocity is very close to zero (and not shown in the plot).
Hence, there is a very rapid transition between different growth behaviors at this finite value of the driving force, and we call the regime $1<\Delta<2.6$ the ``creep branch''.
Notice that the literal Griffith point is located at $\Delta=1$, but nevertheless it seems that significant crack growth starts only at a substantially higher driving force (``apparent Griffith threshold'').
In the creep branch almost all elastic energy is dissipated by viscoelastic damping;
for a more detailed discussion of this point we refer to \cite{SpatschekBrenerPilipenko2008}.

Fig.~\ref{ViscoSDheight} shows the crack height $h/h_0$ with $h_0=(D_s\tau)^{1/4}$ as function of the driving force for surface diffusion in the viscoelastic limit.
Again, the results are obtained by the multipole expansion method.
The behavior is similar as for the crack velocity, as we also see here a monotonic increase as function of the driving force, since this increases the energy dissipation at the crack surfaces.
When the driving force is reduced, the crack tip scale suddenly becomes very small at $\Delta\approx 2.6$, and below this value the crack becomes very sharp, $h/h_0\ll 1$.

\begin{figure}
\begin{center}
\includegraphics[width=1\linewidth]{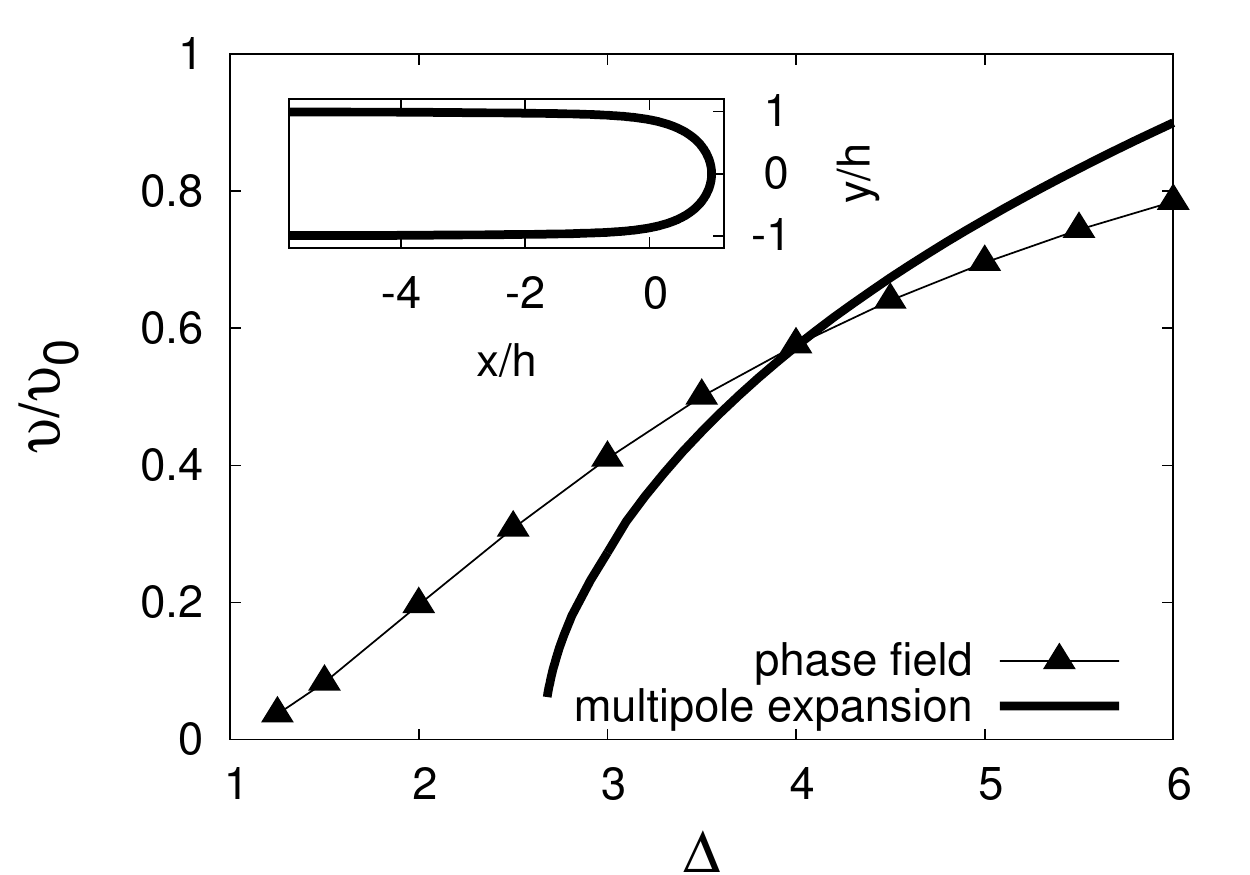}
\caption{Comparison of the tip velocity $\velo/\velo_{0}$ as
a function of the driving force $\Delta$ for mode I fracture using
the PT model, with $\velo_{0}=({D/\tau})^{1/2}$. The solid line corresponds to the results of the multipole
expansion method with infinite viscosity strength. The triangles correspond
to the phase field results with viscosity strength $\chi=2$. The
inset shows the crack shape for the PT model with $\Delta=3.6$ obtained
with the multipole expansion method. Both directions $x$ and $y$
are scaled with the half tail opening $h$ of the crack. }
\label{mode1_visco} 
\end{center}
\end{figure}

Next, we discuss the results to the same regime of slow mode I crack growth (viscoelastic regime), but with the PT mechanism.
Here, we performed simulations using both the multipole expansion approach and phase field modeling.
We point out that the phase field model does not contain only viscoelastic damping but also inertial effects, i.e.~the appearance of the acceleration term in the Newtonian equations of motion.
For the phase field results in Fig.~\ref{mode1_visco} we use $\chi=2$, thus $\velo_R=\sqrt{2} \velo_0$.
This means that the velocity scales are not fully separated, and as soon as $\velo \approx \velo_0$, the velocity has already reached a substantial fraction of the the sound speed, and dynamical effects start to become relevant.
Since the crack speed is ultimately limited by the Rayleigh speed, it is therefore not surprising that the velocities obtained by the phase field method are lower than those by the multipole expansion technique, which assumes $v/v_R\ll 1$.
This behavior is visible in Fig.~\ref{mode1_visco} for driving forces $\Delta>4$.

\begin{figure}
\begin{center}
a)\includegraphics[width=0.8\linewidth]{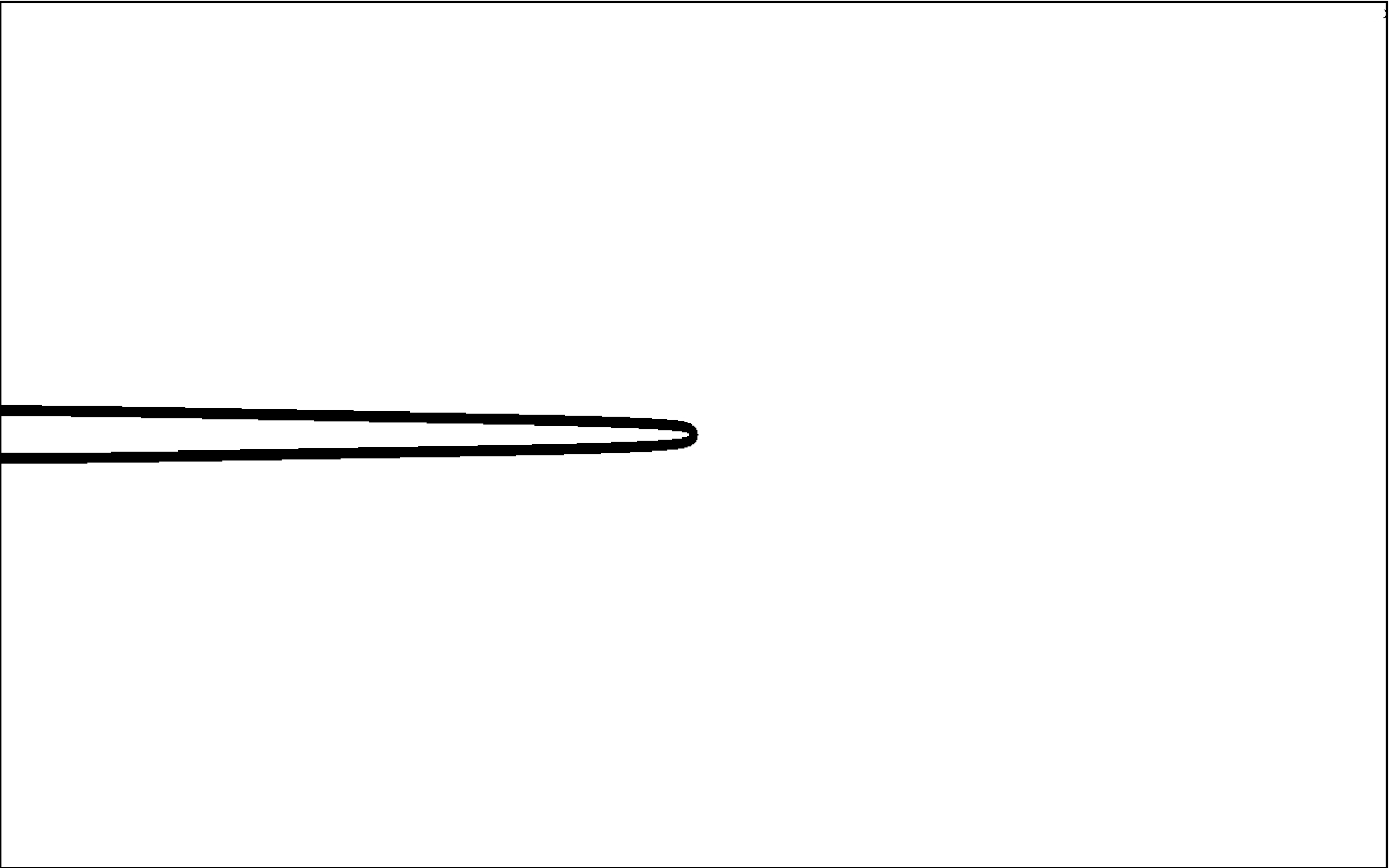}~\\
b)\includegraphics[width=0.8\linewidth]{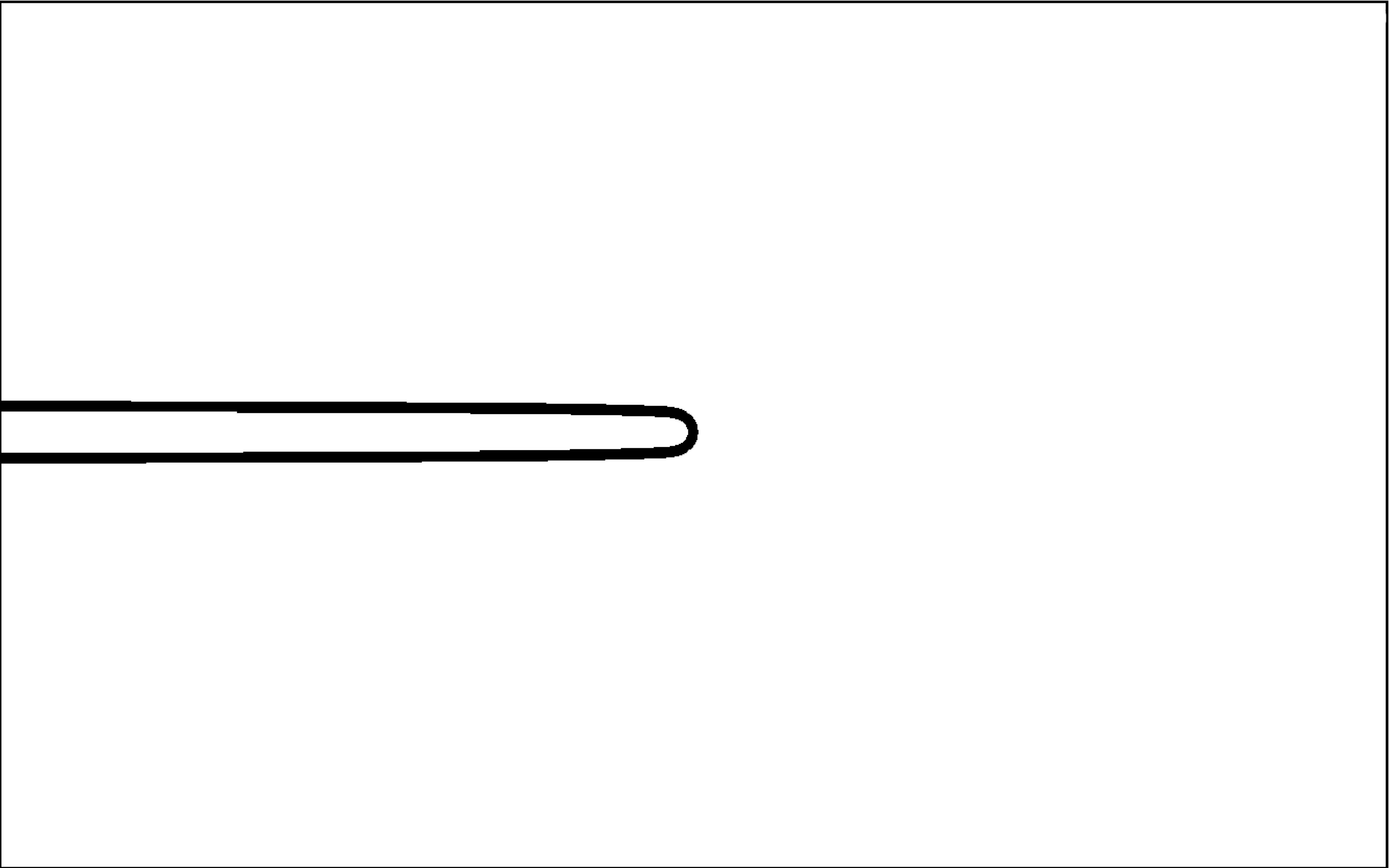}~ \\
c)\includegraphics[width=0.8\linewidth]{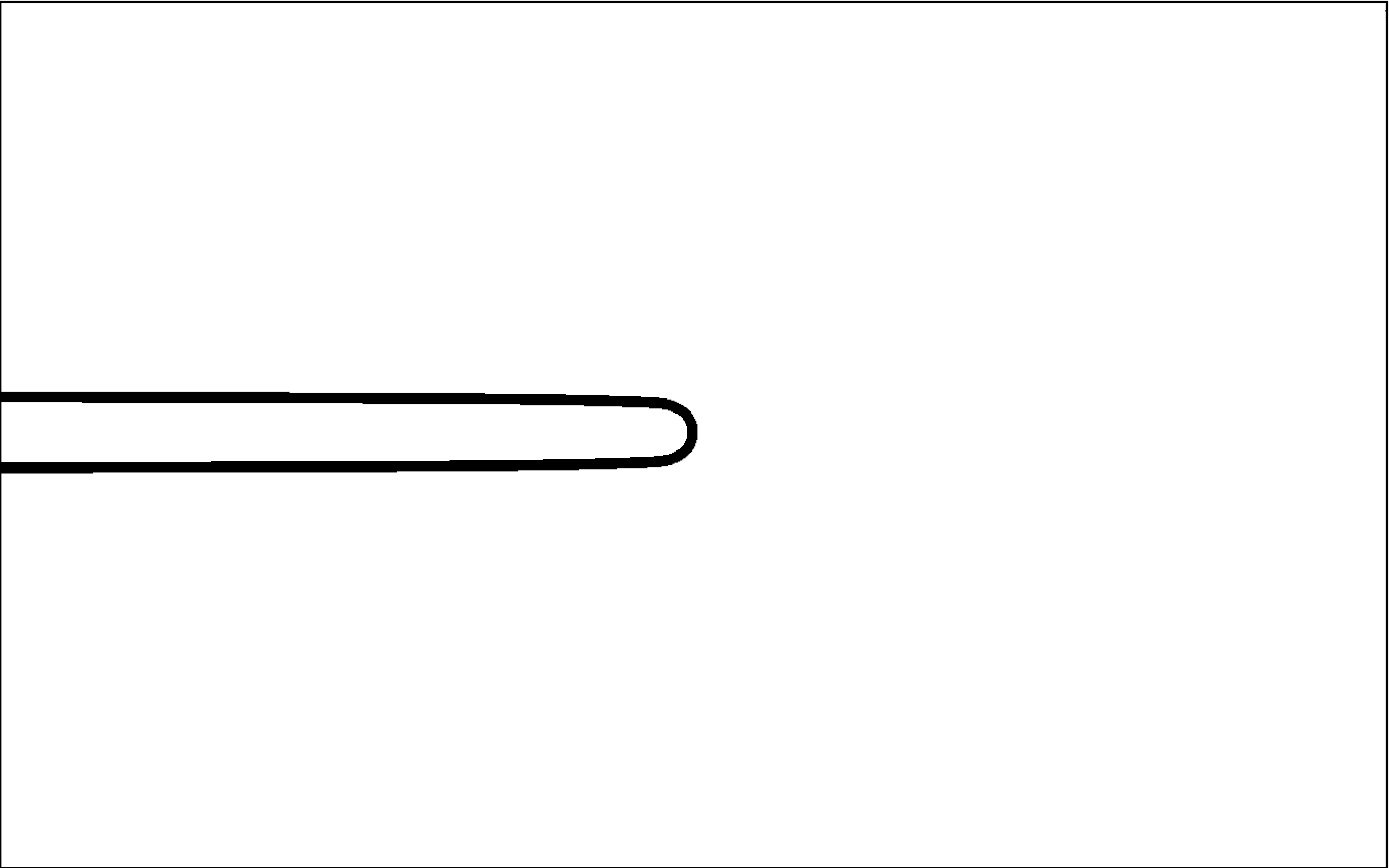}
\caption{ Phase field crack shapes for $\chi=2$
and different driving forces after the time $t/\tau=24.4$:  a) $\Delta=1.25$ b) $\Delta=3.6$ and c) $\Delta=5.0$. We set $W/h_0=60.9$ and $h_0/\xi=2.6$. The thickness of the
interface corresponds the phase field interface width. For high driving forces, the tip radius does not depend on the interface thickness. Notice that for the lowest driving force the crack opening is not constant along the crack but increases towards the tail. The reason is that due to the elastic energy stored in the strip there is an effective short-range repulsion between free surfaces.
}
\label{steady_state_shapefig} 
\end{center}
\end{figure}

Overall, the behavior for the PT model is very similar to the SD results, which includes in particular a monotonic increase of the steady state velocity as function of the driving force.
Again, we did not find indications for crack branching at higher velocities without inertial effects, $\velo\ll \velo_R$.
As for the SD model, the solution branch with $\velo\sim\velo_0$ terminates at $\Delta\approx 2.6$, and below we have again a creep branch with $\velo \ll \velo_0$.
Similarly, we also have here the drop of the crack tip scale $h/h_0$ (with $h_0=(D\tau)^{1/2}$ for the PT mechanism) to very small values for $\Delta<2.6$.
Here, the phase field model brings in another effect, which comes from the interface thickness $\xi$ as intrinsic numerical parameter.
To obtain results that coincide with the multipole expansion method it is necessary to maintain the scale separation $\xi/h \ll 1$.
We have demonstrated for the inertial regime that it is possible to reach this limit, although it is numerically very demanding \cite{Spatschek2007}.
Here, however, we rather consider the interface thickness as additional ``microscopic'' cutoff scale, which prevents that the crack tip scale drops below this value;
this effect can be seen from the steady state crack shapes, see Fig.~\ref{steady_state_shapefig}.
Therefore, for phase field modeling the creep branch does not exist, and consequently the crack velocity continues to decay smoothly down to $\Delta=1$.

For moderate driving forces around $\Delta=4$, the qualitative agreement
between phase field results and the velocities from the multipole
expansion method is good.

For comparison with the phase field shapes the inset of Fig.~\ref{mode1_visco} shows a typical
steady state crack contour obtained with the multipole expansion method
in the limit of static viscoelasticity.
Again, the shape is drawn without elastic
displacements, which should be added to obtain the real shape under
load. The crack tip
scale is selected self-consistently, and the finite time cusp singularity
of the ATG instability does not occur. Therefore, we can conclude
that the sole presence of viscous bulk dissipation is a way
to cure this well-known problem \cite{SpatschekBrenerPilipenko2008}.
Since we do not have a conservation law for the amount of material inside the crack, the pattern looks different in comparison to the SD crack shape.

Finally, we remark, that within the ``static'' limit (without inertial effects) a branching
instability does not show up for mode I fracture neither in the PT model nor for the SD mechanism.
The phase field model of course contains inertial effects, and therefore for $\Delta\gg 1$ one always find branching events once $\velo\sim \velo_R$.
In contrast, for pure mode III fracture and mixed mode scenarios within the SD model, which will be discussed below, an instability appears even in the static limit \cite{SpatschekBrenerPilipenko2008}.

\subsubsection{Inertial limit}
\label{Mode1_pure_dynamics} %

Here we consider situations where the crack speed becomes comparable to the Rayleigh speed, while it is assumed that the viscous damping is negligible, i.e.~$\chi=0$.
Surprisingly, for SD no physically reasonable steady state solutions exist, and therefore we discuss only the PT mechanism.
For that, we briefly review the results of our previous work \cite{Spatchek2006,pilipenko2007,Spatschek2007}.

\begin{figure}
\begin{center}
\includegraphics[width=1\linewidth]{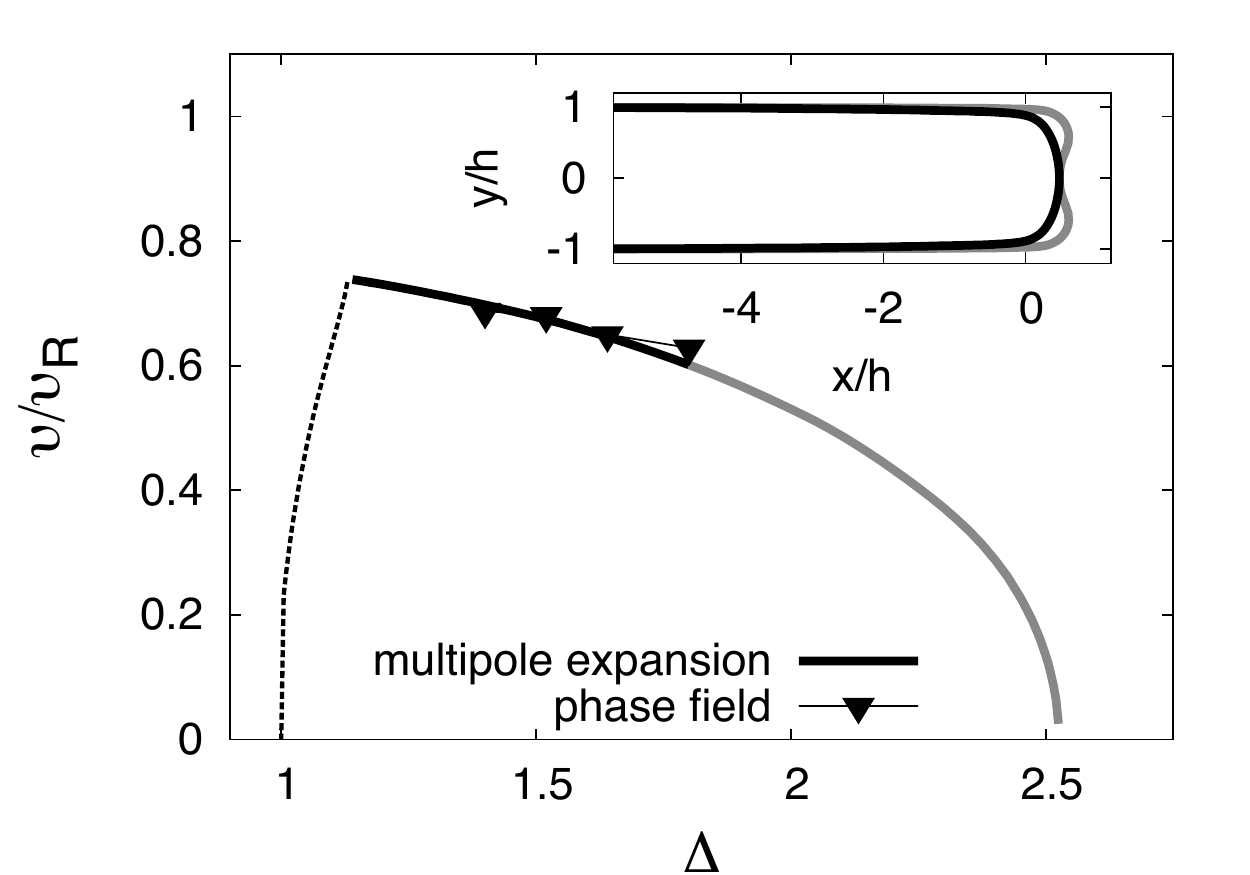}
\caption{Quantitative comparison of steady-state crack velocities obtained from the multipole expansion technique and the extrapolated values from phase field simulations, for the limiting case of $\chi=0$.
The gray line-color indicates results where a negative tip curvature was measured. 
The inset of the figure shows multipole expansion crack shapes of the stable (black curve $\Delta=1.3$) and the unstable solution (gray curve $\Delta=2.3$). 
Both directions $x$ and $y$ are scaled with the half tail opening $h$ of the crack.
Below the point $\Delta_c\sim1.14$, indicated by the dotted line, we show the velocity of the dissipation-free solution, where the tip radius $r_0$ is selected by a microscopic length scale.
}
\label{VelocityComparison}
\end{center}
\end{figure}

Here, rather small scale phase field simulations \cite{Spatchek2006} delivered a picture, which was in conflict with very precise multipole expansion method results \cite{pilipenko2007}, since the predicted driving force dependence of the steady state velocity came out with opposite slope.
This discrepancy was resolved in \cite{Spatschek2007} by performing a large series of phase field simulations for different system sizes $W$ and tip scales, and careful extrapolation of the crack velocity to the limit $\xi/h\to 0$ and $h/W\to 0$.
The main result in this context was the quantitative agreement of the steady state crack velocities $\velo/\velo_{R}$ obtained from both numerical methods, as shown in Fig.~\ref{VelocityComparison}.
The small deviation for $\Delta=1.8$ is due to the fact that this value is already close to the threshold of the branching instability, which cannot be captured by the multipole expansion method.
With this costly quantitative comparison, we found in particular evidence
for the remarkable prediction that the steady state velocity decays
weakly with increasing driving force \cite{pilipenko2007}.
Nevertheless, the product $\velo h/D$, which controls the dissipation, is
still growing monotonically. This counterintuitive outcome
means that within the dynamic limit ($\chi=0$) of the model
the dissipation is mainly increased due to tip blunting instead of
a rise of the crack speed.
Tip blunting then always leads to a tip branching instability for higher driving forces, due to a secondary ATG-instability as mentioned in section \ref{selection}. 
In the multipole expansion method, which captures only steady state solutions, this transition towards unstable crack growth is reflected by a change of sign of the tip curvature at $\Delta\approx 1.8$, which is in agreement with the critical driving force for the branching instability in phase field modeling.
In Fig.~\ref{VelocityComparison} we indicate this change by a change from the black to the gray line-color, and in the inset we show two corresponding crack shapes.
For further details we refer to \cite{pilipenko2007}.

Although the model provides a selection of the crack tip scale and velocity in the limit of vanishing dissipation, it suffers from the fact that for the small range of driving forces near the Griffith point the velocity of the crack is finite while the size of the crack tip approaches zero.
More precisely, the velocity branch in Fig.~\ref{VelocityComparison} terminates at $\Delta=1.14$ with a finite velocity, and below this value no steady state solutions are found.
Thus, the solution branch does not naturally connect to the Griffith point $\Delta=1$.
Here, another tip scale mechanism would be necessary, in order to restore selection.
In the phase field method (without performing the extrapolation to the sharp interface limit $\xi/h\to 0$) the interface thickness serves here as numerical tip scale selection mechanism \cite{Spatchek2006}.
In the same way, setting the minimal allowed opening by hand, also the situation for the multipole expansion method can be improved \cite{pilipenko2007}. However, since our intention was to formulate a continuum model of fracture which is
independent of any microscopic length scales, for the present purpose,
such introduction of a finite cutoff length scale is unsatisfactory.
Then, on the other hand, the sudden velocity jump and the subsequent decay with increasing driving force are unavoidable outcomes and seem to disagree with intuitive expectations.
This, together with the fact, that in the inertial limit no steady state solutions exist for SD, has been a major motivation for considering viscoelastic bulk dissipation as an alternative selection mechanism.
We point out that the selection of tip velocity and radius via viscous bulk dissipation as discussed in the previous subsection
neither suffers from the problem of finite velocities slightly above
the Griffith point nor requires the introduction of an additional
microscopic cutoff length scale. On the other hand, for mode I the
tip branching instability does not occur without inertial effects.
Therefore, to describe the full picture of crack
propagation under mode I loading, it is highly desirable to account
for both viscous dissipation as well as dynamic effects, as will be discussed in the next subsection.

\subsubsection{Viscoelastodynamic regime}

The rigorous treatment of the regime where both inertial and viscoelastic effects are relevant has been performed with the phase field model, and additionally an approximative description with the multipole expansion method is possible.
The model, which has been introduced in section \ref{toy_model}, has the advantage that it is exact without viscous damping in the inertial regime, $\chi=0$, and it is a rigorous perturbation expansion for slow cracks, $v\ll v_R$, using the viscosity $v\tau/h$ as small expansion parameter, where $h$ is the crack tip scale, 
hence operating in the regime of large values of $\chi\to\infty$.
Therefore, this model allows to gain qualitative insights into the full problem of dynamic mode I crack propagation including
viscous bulk friction, both for the PT mechanism and SD.

\begin{figure}
\begin{center}
\includegraphics[width=1\linewidth]{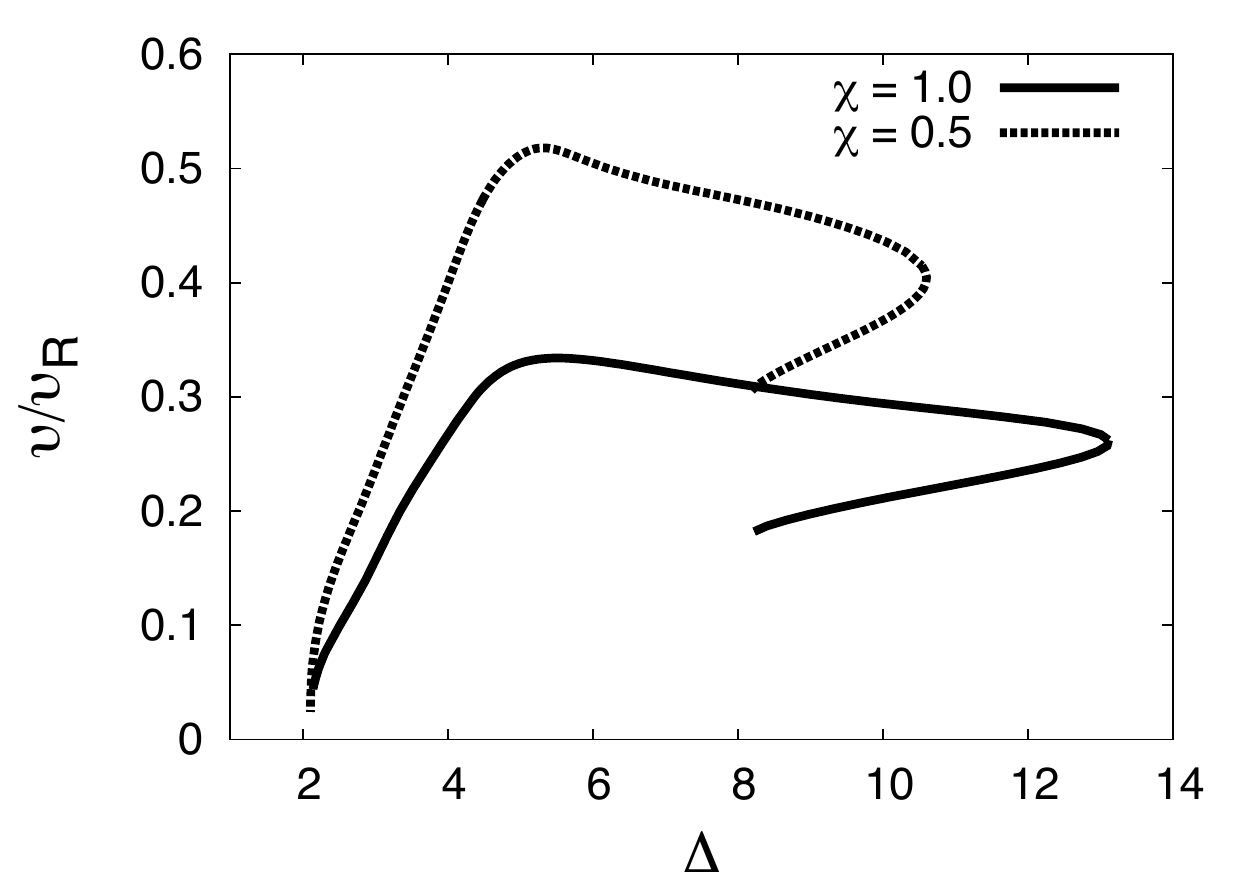} 
\caption{Viscoelastodynamic model results for the steady state velocity $\velo/\velo_{R}$
as a function of the driving force $\Delta$ in case of mode I crack
propagation using the SD model. The solid line corresponds to $\chi=1.0$,
the dashed one to $\chi=0.5$. }
\label{SDvelocity} 
\end{center}
\end{figure}

We start the discussion of the results again for the SD model, and the results are all obtained by the multipole expansion method.
As mentioned above, no physically reasonable steady state solutions exist in the purely inertial limit, $\chi=0$, and we will return to this point below.
However, with the inclusion of viscous effects within the present model, steady state solutions exist for finite values of $\chi$, as shown in Fig.~\ref{SDvelocity}.
In this plot the velocity is shown on the scale of the Rayleigh speed $v_R$, which equals the viscous scale $v_0$ for $\chi=1$.
As in the purely viscous limit, fast growth does not start at the literal Griffith point $\Delta=1$, but a higher value, which is located at around $\Delta\approx 2.1$ for both shown paremeters $\chi=0.5$ and $\chi=1$
(below this value we have again the creep branch with $\velo\ll \velo_0$). 
From this point on the velocity first increases with increasing driving force until it reaches a maximum and then it again decreases until the termination in a bifurcation.
We expect crack branching beyond this point, since steady state solutions do not exist in this regime.

In agreement with the intuitive expectation that the cracks should slow down by viscous damping, the velocity decreases with the increase of the viscosity strength $\chi=\velo_{R}^{2}/\sqrt{D_{s}\tau^{-3}}$.
At the same time, the bifurcation point, beyond which no steady state solutions exist, is shifted to higher driving forces.
This fact indicates that viscous damping suppresses crack branching, in agreement with the earlier observation that it does not occur in the purely viscous limit.
For very strong viscous damping, $\chi\to\infty$, the velocity becomes small and is then set by the viscous speed scale $v_0$, as depicted in Fig.~\ref{mode1_visco_SD}, where it is a purely monotonically growing function of the driving force.

On the other hand, the results show that upon reduction of the viscous strength, i.e.~for smaller values of $\chi$ the velocity increases.
Finally, the curves first touch and then terminate at the Rayleigh velocity, which is the upper theoretical limit for mode I fracture.
Then, for the inertia limit the curves would start with the decaying part of the curve at a finite value of $\Delta>1$ but with $v=v_R$, and we do not consider this as a physically plausible solution.
However, in the framework of the present model it becomes thereby understandable why no ``reasonable'' solutions exist in the elastodynamic limit.

\begin{figure}
\begin{center}
\includegraphics[width=1\linewidth]{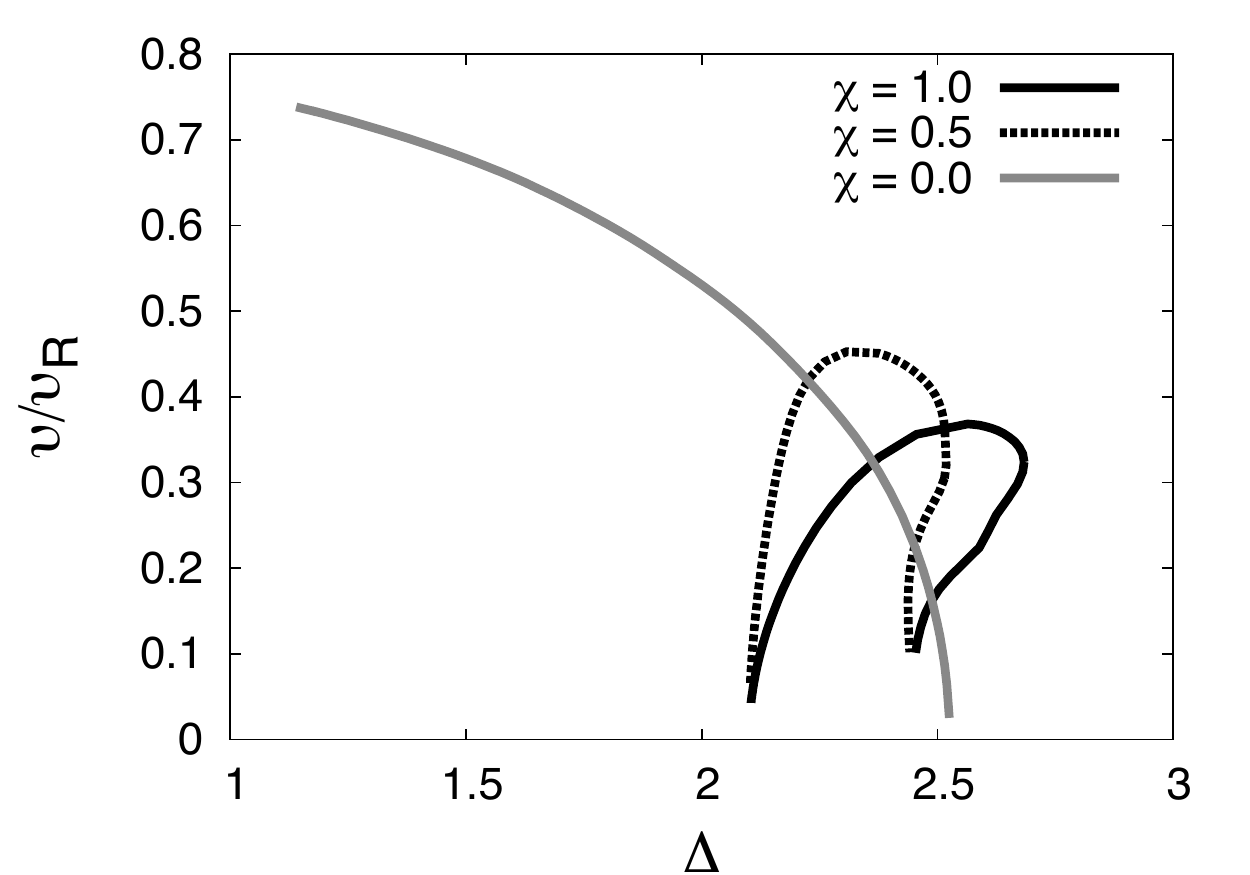} 
\caption{Viscoelastodynamic model results for the steady state tip velocity $\velo/\velo_{R}$
as a function of the driving force $\Delta$ in case of mode I crack
propagation using the PT model. The solid black line corresponds to $\chi=1.0$,
the dashed to $\chi=0.5$. The gray colored solid line shows the steady state velocities in the inertial limit, when $\chi=0.0$ (see Fig.~\ref{VelocityComparison}).
}
\label{PTvelocity}
\end{center}
\end{figure}

Next, we inspect the behavior of the model for PT dynamics, as obtained from the multipole expansion method.
The results for the same two different values of $\chi$ are shown in Fig.~\ref{PTvelocity}, and exhibit a qualitatively similar behavior as for SD.
Also here, the growth starts from an ``apparent'' Griffith point, which coincides with the value for SD, and below we find ``creep solutions'' with very low velocities, which are not shown in the plot.
The reason for the agreement of these apparent Griffith points is that for $v\ll v_0$ the chemical potential is basically zero along the crack, both for SD and PT;
also, below this value of the driving force the behavior is dictated by bulk dissipation, and therefore the transport mechanism on the crack surfaces plays only a minor role.

From the apparent Griffith point on the velocities increase monotonically quite rapidly up to a maximal value. 
Then the velocity maximum is followed by a small range of driving forces, where the velocity decreases with increasing driving force. 
Crack branching is expected beyond the bifurcation point.
With increasing viscosity strength
$\chi$ the driving force of maximal velocity as well as the
point where the curvature turns negative are both shifted to higher
driving forces. 
This supports the conclusion that dynamic effects favor of the occurrence the tip splitting instability, or -- vice versa -- the presence of viscous bulk friction helps to stabilize the crack against the tip splitting instability, which is also qualitatively supported by fully dynamic phase field simulations.

For $\chi\rightarrow\infty$ the results come closer to the previous curve for the viscoelastic limit of the PT model, as shown in Fig.~\ref{mode1_visco}, where the velocity is then on the scale $v_0$.
In contrast, with a
decrease of the viscosity strength the point of maximal velocity is
shifted more and more to the left until, finally, in the case of vanishing
viscosity, $\chi=0$, only the decaying part remains (see also Fig.~\ref{VelocityComparison}).

The PT and SD model have in common that the crack velocity decreases with the increase of viscosity strength $\chi$, while the turning point is shifted to higher driving forces.
A difference is, that the point of maximal velocity and moreover
the tip curvature turning point appear at much higher driving forces for the SD 
than for the PT model, which especially also enlarges
the regime, where the velocity decreases with increasing driving force. 

\begin{figure}
\begin{center}
\includegraphics[width=0.45\linewidth]{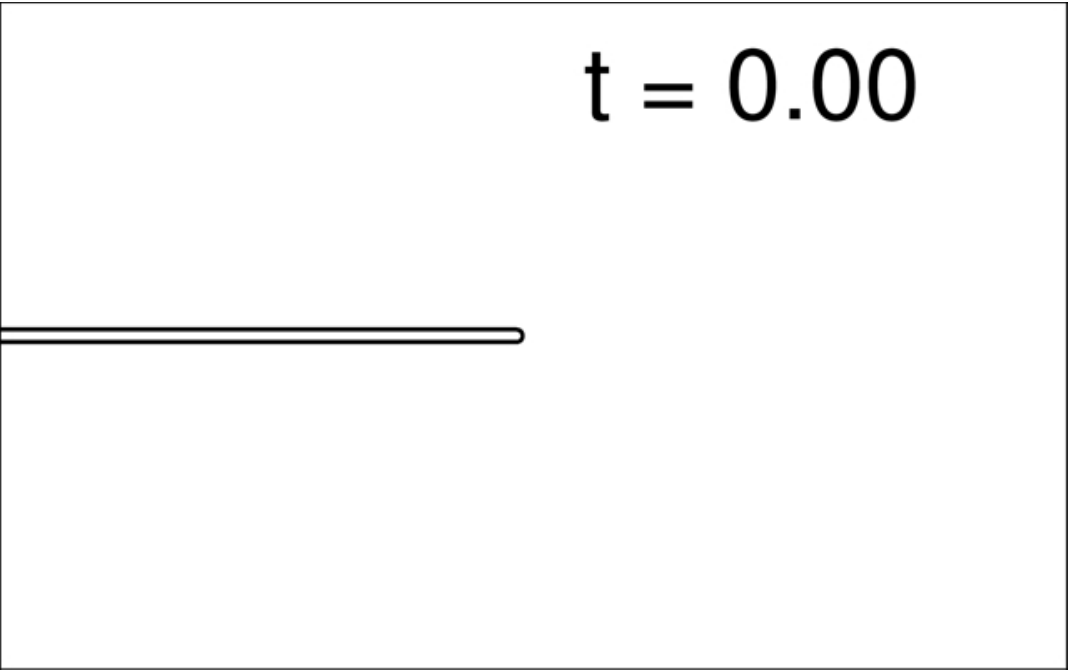}
\includegraphics[width=0.45\linewidth]{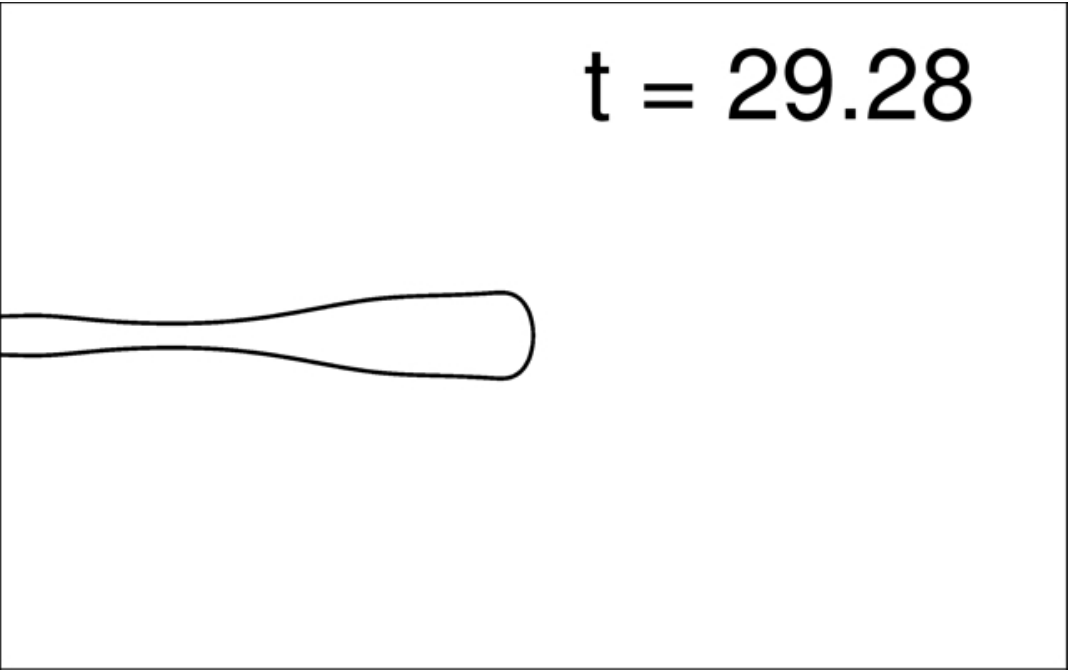}
\includegraphics[width=0.45\linewidth]{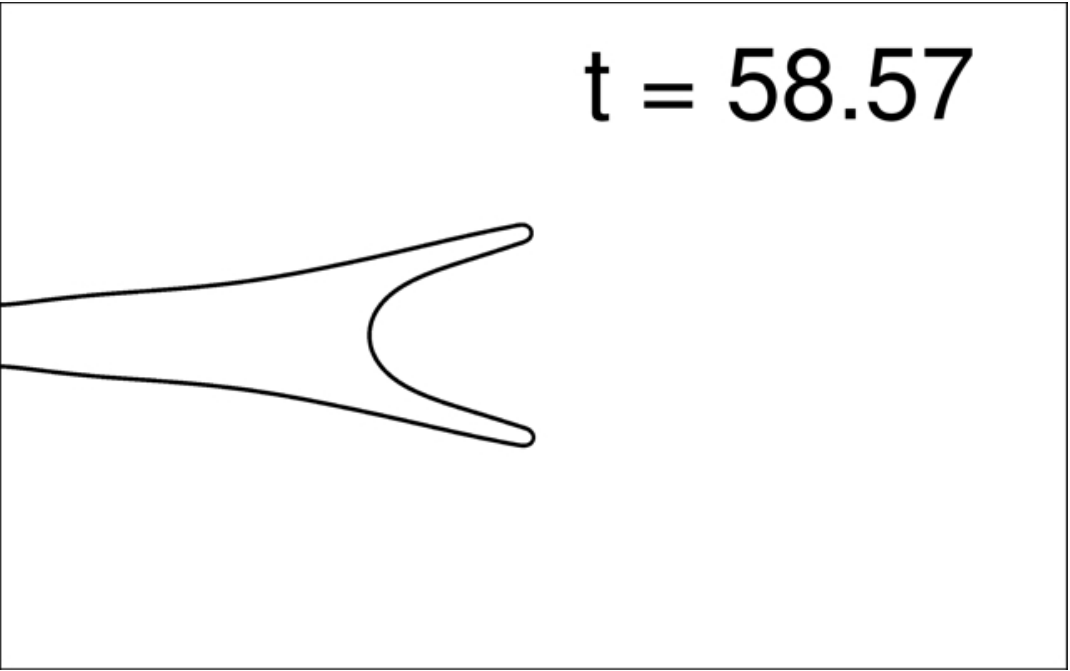}
\includegraphics[width=0.45\linewidth]{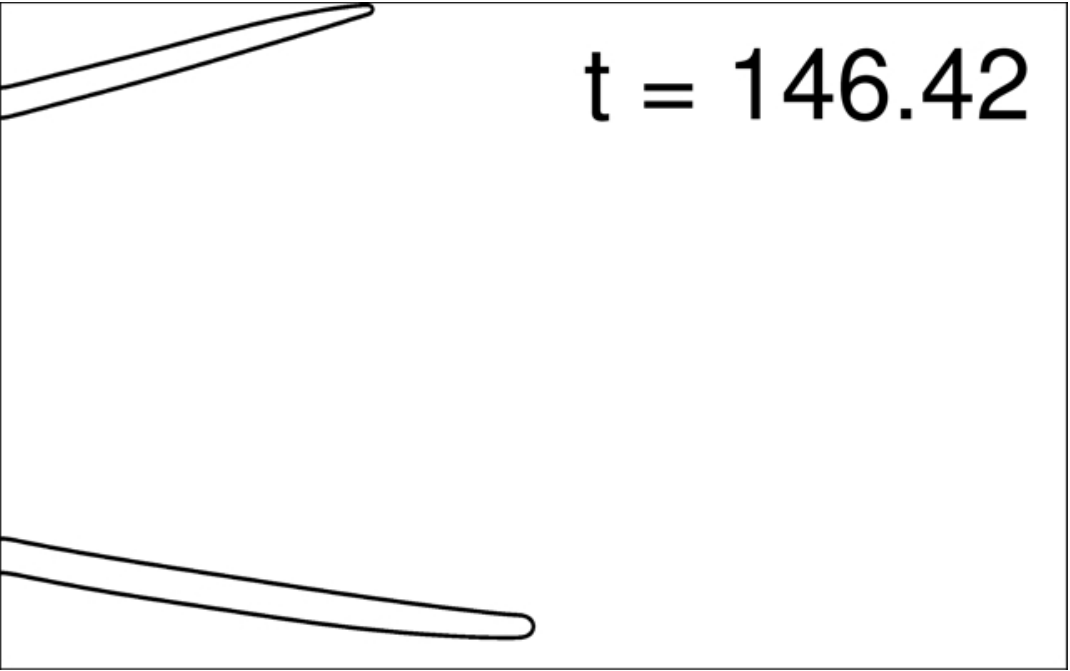}
\includegraphics[width=0.45\linewidth]{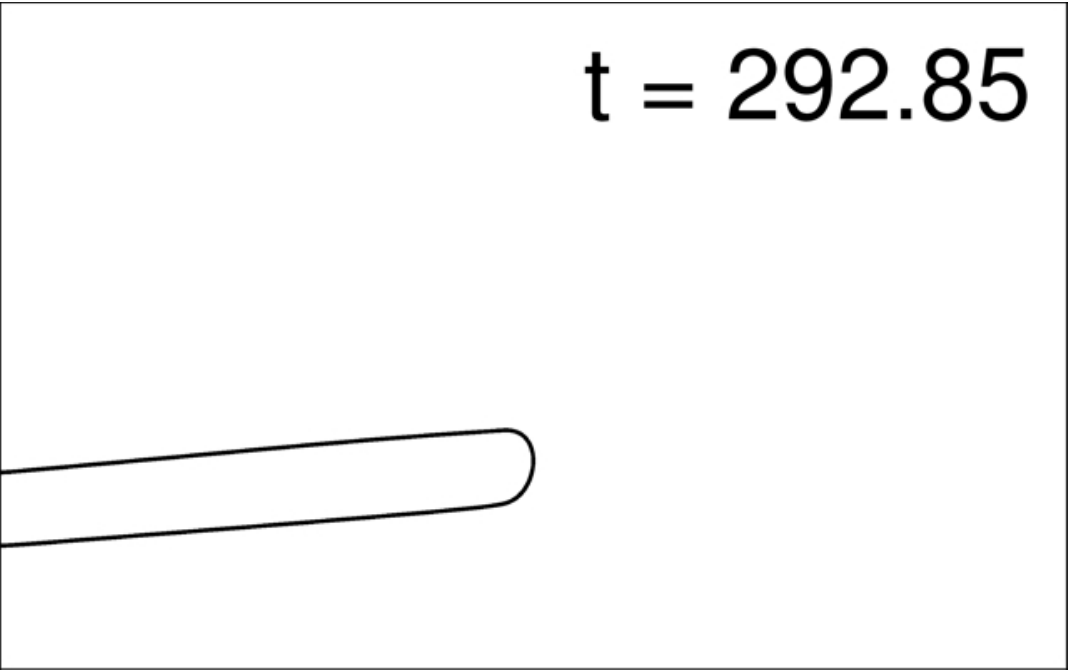}
\includegraphics[width=0.45\linewidth]{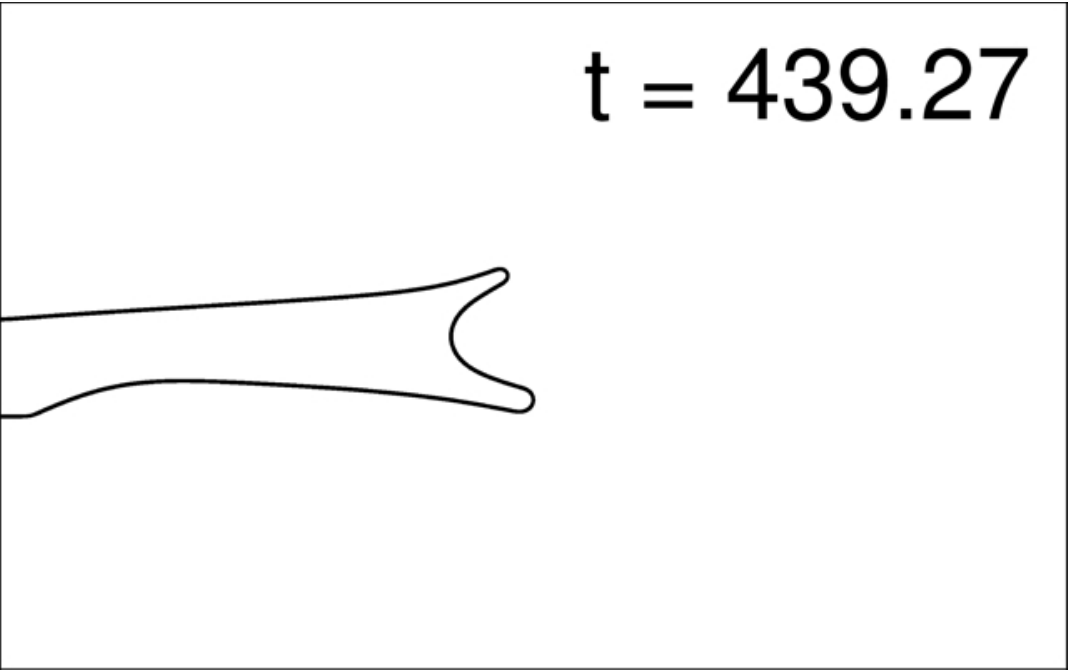} 
\caption{ Irregular tip splitting scenario for a viscosity strength of $\chi=2$
and $\Delta=10.0$. We set $W\velo_{R}/D=170$ and $D/\velo_{R}\xi=1.9$,
and the size of the system is $1600\times4096$ grid points. The
time $t$ is given in units $D/\velo_{R}^{2}$. The thickness of the
interface corresponds the phase field interface width. }
\label{splitfig} 
\end{center}
\end{figure}

The phase field method allows to study crack growth also in regimes, where no steady state solutions exist.
However, a quantitative determination of the onset of the branching instability is computationally very expensive. 
The onset of the irregular tip splitting behavior depends, in particular, sensitively on the system size, because in relatively small systems the branches of the crack cannot separate since they are repelled by the boundaries. 
Therefore, the steady state growth is always stabilized by finite size effects. 
On the other hand, initial conditions can trigger an instability, and then a long transient is
required to get back to steady state solutions. 
However, as shown in Fig.~\ref{splitfig}, even for a relatively high viscosity strength $\chi=2$ we find the irregular tip splitting behavior for $\Delta=10.0$.

\subsection{Mixed Mode fracture}
\label{mode3}

Here we discuss predictions of the model beyond a pure mode I loading.
It turns out that the results change even qualitatively for mode III.
For a broader picture, we study also situations with mixed mode loading (mode I + mode III), but still assume that the crack shape is translational invariant in the direction of the crack front line.
We therefore suppress effects like the development of a helical instability as studied in \cite{PonsKarma}.
We limit our investigations here to viscous effects (i.e. $v\ll v_R$), and only for the phase field model we also take into account inertial effects.
As before, we make the simplifying assumption $\nu=\zeta=1/3$, and therefore viscosity introduces only one additional time scale, $\tau=\eta/E$.

\begin{figure}
\begin{center}
\includegraphics[width=1\linewidth]{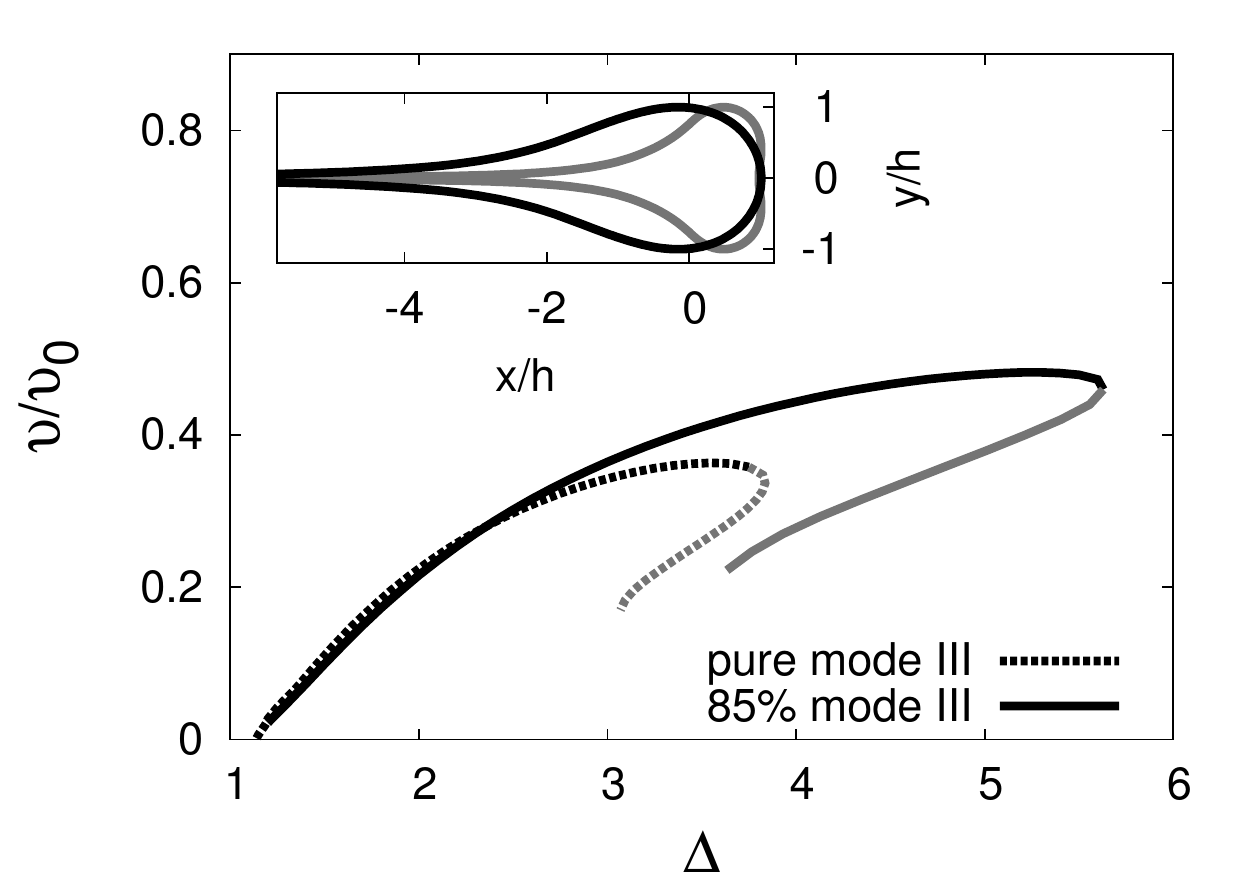}
\caption{Steady state propagation velocity as function of the driving force for pure mode III and a mixture with $\Delta_{III}/\Delta=0.85$ are displayed, for the SD model. 
The gray line belongs to steady state solutions with negative tip curvature. 
The inset of the figure shows the steady state crack shapes of the
stable and the unstable solution in the case of mixed mode loading
with $\Delta_{III}/\Delta=0.85$ and a total driving force of $\Delta=3.6$.
Both directions $x$ and $y$ are scaled with the half maximum height
$h$ of the crack. 
}
\label{MixedModeVelocitiesSD} 
\end{center}
\end{figure}

First, we review our findings concerning the crack behavior
in the mixed mode scenario for the case of the SD model \cite{SpatschekBrenerPilipenko2008}, which were obtained by the multipole expansion method.
As shown in Fig.~\ref{MixedModeVelocitiesSD}, 
the crack speed increases with the driving force for pure mode III, until it reaches
a maximum at $\Delta\approx3.5$, then it decreases, and obviously
steady state solutions do not exist beyond the point $\Delta\approx3.8$,
where the stable branch merges with another (unstable) solution. 
Beyond
the bifurcation point $\Delta\approx3.8$ we expect crack branching,
in analogy to our findings for fast dynamic mode I fracture, as discussed
in the previous section. It is quite remarkable, that the presence
of mode III loading contribution leads to the occurrence of the tip
branching instability even in the case of static elasticity.

\begin{figure}
\begin{center}
\includegraphics[width=1\linewidth]{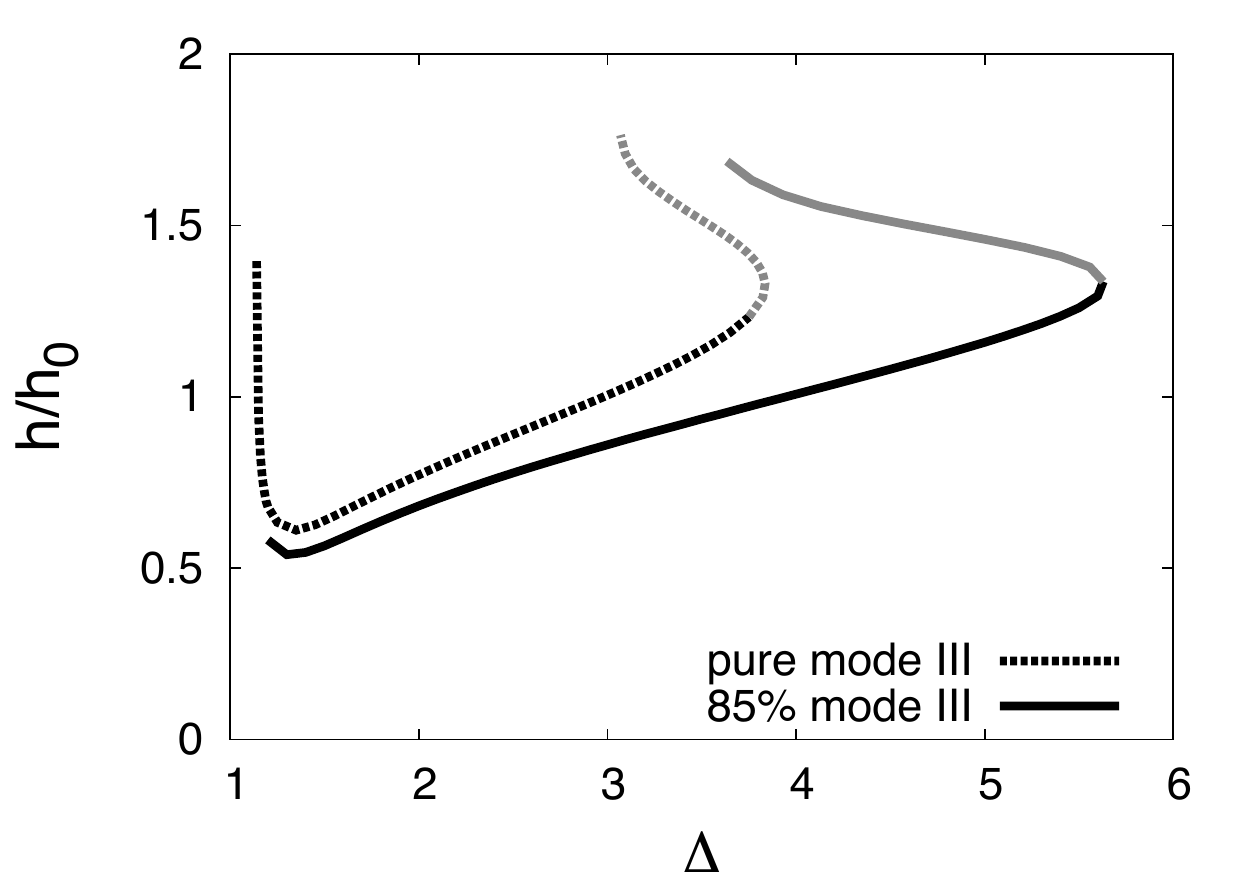}

\caption{Half crack height as function of the driving force for pure mode III and a mixture with $\Delta_{III}/\Delta=0.85$, for the SD model. The lengthscale used here is $h_0=(D\tau)^{1/4}$. 
The gray line belongs to steady state solutions with negative tip curvature. 
}
\label{fig3}
\end{center}
\end{figure}

In Fig.~\ref{fig3} we also show the maximum height of the crack as function of the driving
force for different loadings. 
At $\Delta\approx1.1$ the size of the mode III steady state crack diverges and $\velo\to0$. 
The viscous dissipation becomes negligible here, but the surface dissipation remains finite. 
This point can be interpreted as the point of ductile-to-brittle transition: Below it the size grows indefinitely in time and the crack slows down, while above this point steady state solutions with a finite tip scale exist.

Starting from a pure mode III crack, we can now include additional mode I loadings. 
Fig.~\ref{MixedModeVelocitiesSD} shows that this shifts the bifurcation point towards higher values and therefore extends the range of steady state solutions towards higher driving forces.
From this we can conclude that mode III contributions favor the appearance of the tip splitting instability.
In contrast, the preceding results suggets that inertial effects should push the onset of branching back towards lower driving forces.

It is important to note that mode I and mode III have a different behavior, which is due to the behavior of the stresses on the crack surfaces.
We focus here on the elastic fields far away from the tip, and in this region the behavior is purely elastic.
Without inertial effects, normal and shear stresses vanish on the crack surfaces, and therefore it is the tangential stress component which determines the elastic contribution to the chemical potential.
From the singular contributions to the stress \cite{Freund:1998yq,Fineberg:1999zr} one obtains on the crack lips elastic chemical potential contributions $\mu_{el}(x\rightarrow-\infty)\sim1/x$ for mode III and $\mu_{el}\sim1/x^{3}$ for mode I.
This weaker decay of the singular fields for mode III also influences the crack shapes.



Let us look at the asymptotic shape $y(x)$ of a crack in the SD model (in the tail region $x\to-\infty$), and focus on polynomial terms.
As discussed in \cite{BrenerSpatch2003} exponential terms are very important for the selection of the crack velocity and tip scale, which is related to the supression of {\em growing} exponentials in the tail region.
Remaining exponentially decaying terms are small in comparison to the power law terms in the asymptotic regime, and therefore we suppress them here.
By the steady state equations of motion Eqs.~(\ref{steady:eq1}) and (\ref{steady:eq3}) we obtain $\velo y(x)\sim D_{s}y'''-const/x^{2}$ for mode III.
Therefore, we obtain a scaling behavior for surface diffusion as $y(x)\sim x^{-2}$.
Correspondingly, in the case of pure mode I loading for SD
the shape function decays like $y(x)\sim x^{-4}$, which is substantially
faster. 

\begin{figure}
\begin{center}
\includegraphics[width=1\linewidth]{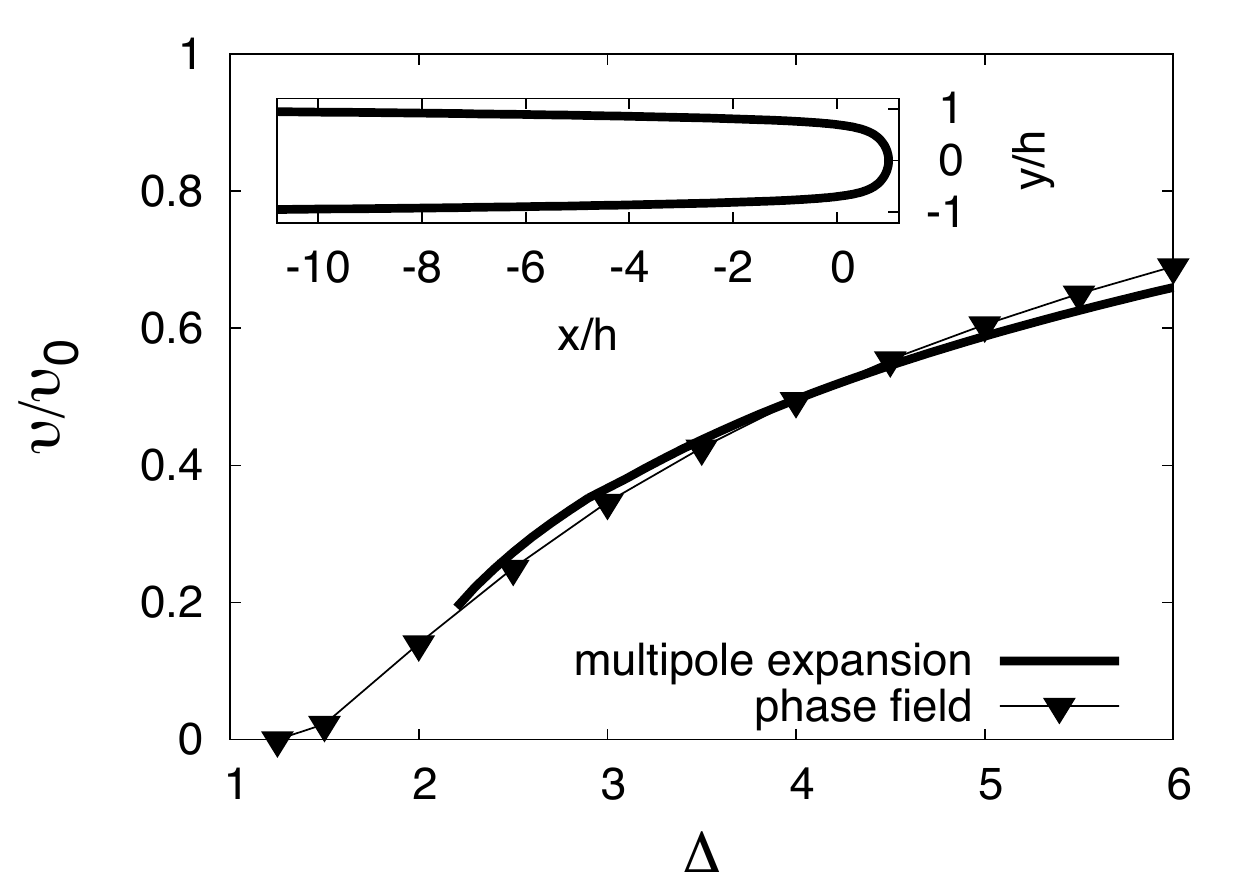} 
\caption{Qualitative comparison of crack tip velocities $\velo/\velo_{0}$
as a function of driving force $\Delta$ for $50\%$ mode III loading
in the case of the PT model. The solid line corresponds to the results
of the multipole expansion method in the viscous limit.
The symbols correspond to the phase field results with a viscosity
strength $\chi=2$. 
For the phase field, we used $W v_R/D=86$ and $D/v_R \xi=1.9$; the size of the system in grid points is $2048\times800$.
The inset shows the multipole expansion method steady state crack shape for a total driving force of $\Delta=12.0$. Both directions $x$ and $y$ are scaled with the half tail-opening $h$ of the crack. 
}
\label{mode3_visco} 
\end{center}
\end{figure}

In the case of the PT model this effect is more pronounced.
Since the amount of ``material'' inside the crack is not conserved, it is reasonable to look for shape functions $y(x) = h + \delta y(x)$, again with purely polynomial functions $\delta y(x)$.
Neglecting the second derivative $\delta y''(x)$ from the curvature contribution to the chemical potential, we obtain from Eqs.~(\ref{chemical_pot_PT})
and (\ref{steady:eq2}) in the co-moving frame of reference the
following ordinary differential equation in the asymptotic regime:
$-\velo\delta y'(x)\sim1/x$. Hence, in the case of a finite mode III
contribution the shape function does not even decay but instead weakly
grows like $y(x)\sim\ln(x)$. 
This slow opening of the crack becomes negligible for higher crack speeds.
This weak logarithmic growth of the asymptotic crack shape is also confirmed by the multipole expansion method simulations as shown in the inset of Fig.~\ref{mode3_visco} in the case of $50\%$ mode III contribution. 

For the phase field simulations, which are also shown in Fig.~\ref{mode3_visco} for $\Delta_I/\Delta_{III}=1$ and $\chi=2$, we find a remarkably good agreement with the sharp interface results over the whole range of driving forces.
With the phase field method, we observe two different kinds of growth:
Slightly above the Griffith point up to a driving force of about $\Delta\approx1.4$,
we obtain solutions with almost zero velocity and an asymptotically
growing crack opening similar to what the shape from the multipole
expansion method shows.
Then above this point the solutions seem to regularize, and this weak growth of the shape function is no longer
observable using the phase field method, probably due to the higher growth velocity.
For pure mode III loading this transition point is shifted to an even higher driving force of about $\Delta\approx2.0$, as we analyzed by means of phase field
simulations. 

We point out that these shape peculiarities can also be interpreted from the more general argument that no stationary shapes exist in mode III if only elastic effects are taken into account.
For this loading mode the only two nonvanishing stress components $\sigma_{xz}$ and $\sigma_{yz}$ can be expressed as real and imaginary part of an analytic function.
An equilibrium solution would require a vanishing chemical potential along the entire crack shape, which in turn demands that the aforementioned analytic function is zero there, since the elastic part is quadratic in both stress components.
If an analytical function is zero along a line segment, it must vanish everywhere;
this, however, is not compatible with a nontrivial remote stress field.


\section{Summary and Conclusion}

We have presented a continuum description of fracture in the spirit of elastically driven interfacial pattern formation processes. 
This description leads to moving boundary problems, where not only the propagation velocity but also the entire shape, and especially the tip radius, have to be selected self-consistently. 
In particular, we have discussed two different mechanisms of crack propagation: In the first case the
crack is considered to advance by material diffusion along the crack surface. 
Secondly, we interpret fracture as a first order phase transformation process of the solid material to a ``dense gas'' phase.

Scaling arguments were given that -- to cure the finite time cusp singularity of the ATG instability --
one necessarily needs an independent selection of the tip radius and the steady state velocity, which is not possible if solely static linear elasticity is taken into account. 
Therefore, apart from capillarity and linear elasticity, additional physical effects are required for the determination of additional length and time scales. 
Since we focus on gaining fundamental insights into the phenomenon of fracture, we have concentrated here only on well established theoretical concepts for dynamic elasticity and viscous dissipation.

The arising moving boundary problems have been solved by two complementary
methods. 
First, we have developed an efficient steady state sharp interface method based on the expansion of the mechanical state in eigenfunctions of a straight mathematical cut. 
Second, a fully dynamic phase field description of crack propagation by first order phase transformation processes has been developed. 

With these numerical tools at hand, we have obtained profound insights into the model behavior of our continuum description of fracture. 
In particular, we have extensively discussed mode I fracture, where the coupled influence of dynamic elasticity and viscous dissipation leads to a model behavior which reproduces three important generic features of fracture:
(i) The saturation of steady state velocities appreciably below the Rayleigh speed, (ii) parameter regimes, where the steady state velocity increases with increasing driving force, and (iii) a crack branching instability for high applied tensions. 
Apart from this also mixtures of mode I and mode III loadings have been discussed, and we have found in particular a different behavior of the crack shapes, as well as a change of the branching behavior.
The main results are summarized in Table \ref{table1}.


The future challenge is to combine the ideas, approaches and results from this work to ``conventional'' fracture models, to address the question of the relevance of the different physical mechanisms.
From a thermodynamical point of view, there is a driving force for elastically induced phase transformations, which leads to the fracture models presented here.
However, it has to be addressed to which extend, and under which environmental conditions, these processes can be understood as dominant mechanisms for crack propagation, i.e.~for instance the diffusion along crack surfaces can be an efficient mechanism in comparison to the pure bond breaking.
We note, that due to the small tip scales the material transport is necessary only on very short distances, and therefore a mechanism like surface diffusion, which is usually assumed to be slow, can still lead to fast crack propagation.
Here it should be pointed out that on such small scales a pure continuum description may not be quantitatively accurate, but can still capture the essential physical mechanisms. 
Furthermore, recent experimental investigations of fracture in brittle
gels possibly reveal macroscopic scales \cite{LevineBen-DavFineber032007}.

In general, the question concerning energy barriers should play a central role and should shed light on the relevance of the different mechanisms.
We expect that material transport should become relevant at elevated temperatures.
In the conventional picture for brittle materials a few bonds per atom have to be broken to advance the crack by one lattice unit, and this event takes place very localized at the (sharp) tip.
The energetic cost for such an ``event'' is on the order of $\mathrm{eV/atom}$.
In contrast, for the ``material transport picture'' the overall energetic expense is the same (since the same amount of new interfaces is created), but a change of the bonding situation is required for several atoms.
However, since the diffusing atoms do not have to be completely detached from the surfaces, energetically efficient low-barrier paths may exist for the motion to the next lattice site, and therefore the effective diffusion constant can be relatively high.
Furthermore, surface reactions, as a recently predicted amorphization of diamond, can lead to a bond weakening and material softening of even very brittle materials within short times \cite{Gumbsch} and could facilitate even higher transport rates.

We notice that in some cases nonlinear elastic corrections may play an important role \cite{BouchbinderLivne2008,LivneBouchbinder2008} and even lead to a high-speed oscillatory instability 
\cite{Bouchbinder2009}.
For more ductile materials, plastic processes due to dislocation emission are important, and they have not yet been taken into account.  
We expect bulk dissipation through plasticity to play a similar role as viscous damping, as has been demonstrated above.
Apart from that, these theories introduce the concept of a yield stress $\sigma_y$, which is a natural cutoff for the stress singularity.
Therefore, from point of view of crack tip selection, one can expect the radius $r_0$ to be determined by this cutoff, i.e.~$r_0 \sim K^2/\sigma_y^2$.
Since this leads to strong blunting, surface diffusion is probably not efficient for high driving forces, and in fact the contribution of surface diffusion to the propagation velocity would be a decaying function of the driving force, hence only for small $\Delta$ it may compete with a bond-breaking mechanism together with plastic flow.
Another aspect is that plastic effects lead to large deformations, and therefore from a technical point of view it would then be desirable to describe the material transport processes in the {\em deformed} system, which suggests the use of a Eulerian rather than a Lagrangian description.

\acknowledgments

M.F. and D.P. thank for support by the DFG grant SPP 1418 and R.S. by DFG SPP 1296.
E. B. gratefully acknowledges support by the Erna and Jacob Michael visiting professorship funds at the Weizmann Institute of Science, Rehovot.


\end{document}